\documentclass[ALICE,manyauthors]{cernphprep}
\usepackage[comma,square,numbers,sort&compress]{natbib}
\usepackage{hyperref}
\usepackage{lineno}
\usepackage{color}
\usepackage{caption}
\usepackage{multirow}
\makeatletter{}
\usepackage{float} 

\begin{document}%
\begin{titlepage}
  \PHyear{2017}
  \PHnumber{010}      
  \PHdate{19 January}  
  
  \title{K$^{*}(892)^{0}$ and $\phi(1020)$ meson production at high transverse 
    momentum in pp and Pb--Pb collisions at $\sqrt{s_\mathrm{NN}}$ = 2.76 TeV}
  \ShortTitle{K$^{*}(892)^{0}$ and $\phi(1020)$ meson production at high-$p_{\mathrm{T}}$}   
  \Collaboration{ALICE Collaboration\thanks{See Appendix~\ref{app:collab} for the list of collaboration members}}
  \ShortAuthor{ALICE Collaboration} 
  \begin{abstract}
     The production of K$^{*}(892)^{0}$ and
    $\phi(1020)$ mesons in proton-proton (pp) and lead-lead (Pb--Pb) 
    collisions at $\sqrt{s_\mathrm{NN}} =$ 2.76 TeV has 
    been analyzed using a high luminosity data sample accumulated in 2011 with
    the ALICE detector at the Large Hadron Collider (LHC). Transverse momentum 
    ($p_{\mathrm{T}}$) spectra have been measured for K$^{*}(892)^{0}$ and
    $\phi(1020)$ mesons via their hadronic decay channels for $p_{\mathrm{T}}$ 
    up to 20 GeV/$c$. The measurements in pp collisions have been compared to 
    model calculations and used to determine the nuclear modification factor 
    and particle ratios.  The K$^{*}(892)^{0}$/K ratio exhibits significant 
    reduction from pp to central Pb--Pb collisions, consistent with the 
    suppression of the K$^{*}(892)^{0}$ yield at low $p_{\mathrm{T}}$ due to 
    rescattering of its decay products in the hadronic phase. In central 
    Pb--Pb collisions the $p_{\mathrm{T}}$ dependent $\phi(1020)/\pi$ and 
    K$^{*}(892)^{0}$/$\pi$ ratios show an enhancement over pp collisions for 
    $p_{\mathrm{T}}$ $\sim$3 GeV/$c$, consistent with previous observations of 
    strong radial flow. At high $p_{\mathrm{T}}$, particle ratios in Pb--Pb 
    collisions are similar to those measured in pp collisions. In central 
    Pb--Pb collisions, the production of K$^{*}(892)^{0}$ and $\phi(1020)$ 
    mesons is suppressed for $p_{\mathrm{T}}$ $\textgreater$ 8 GeV/$c$. 
    This suppression is similar to that of charged pions, kaons and protons, 
    indicating that the suppression does not depend on particle mass or flavor
    in the light quark sector. 
    
  \end{abstract}
\end{titlepage}
\setcounter{page}{2}
\section{Introduction}
It has been established that hot and dense strongly interacting matter, 
often described as a strongly-coupled quark-gluon plasma 
(sQGP)~\cite{Adams:2005dq, Adcox:2004mh, Gyulassy:2004zy}, is produced in heavy-ion
collisions at ultrarelativistic energies. The properties of this matter 
are characterized, among others, by the energy loss of partons traversing 
the dense color-charged medium, which manifests itself via suppression 
of hadrons with high transverse momentum in central Pb--Pb collisions.
The hadrons that contain light (up, down and strange) valence quarks exhibit
a similar suppression as particles containing heavy quarks (charm) both at 
RHIC~\cite{Adare:2006ns, Adare:2011yf} and at the LHC~\cite{Adam:2015nna, Adam:2015sza}.
The apparent particle species independence of high-$p_{\mathrm{T}}$ hadron 
suppression is a challenge for models~\cite{Bellwied:2010pr, Liu:2008zb, Liu:2006sf}. Since 
K$^{*}(892)^{0}$ (d$\mathrm{\bar{s}}$), $\mathrm{\overline{K}}^{*}(892)^{0}$
($\mathrm{\bar{d}}$s) and $\phi(1020)$ (s$\mathrm{\bar{s}}$) contain strange 
(or anti-strange) quarks, they are used here for a systematic study of the 
particle species dependence of the partonic energy loss in the medium. 
Moreover, the measurements of high-$p_{\mathrm{T}}$ differential yields can 
be used to test perturbative QCD inspired model calculations.

The system produced in heavy-ion collisions evolves through different stages, 
with a transition from partonic to hadronic matter around a temperature 
T$_{\mathrm{c}}$ $\approx$ 156 MeV~\cite{Floris:2014pta, Borsanyi:2010bp, Aoki:2006br}.
The K$^{*}(892)^{0}$ and $\phi(1020)$ life times in vacuum are 4.16 $\pm$ 0.05 
fm/$c$ and 46.3 $\pm$ 0.4 fm/$c$, respectively~\cite{Agashe:2014kda}. Due to their short 
life times, resonances can be used to probe the system at different timescales 
during its evolution and have been proven to be very useful in exploring various 
aspects of heavy-ion collisions~\cite{Torrieri:2001ue}. Yields of resonances measured 
via hadronic decay channels can be affected by particle rescattering and 
regeneration in the hadron gas phase. The momentum dependence of rescattering 
and regeneration may also modify the observed momentum distributions of the
reconstructed resonances. 

Resonances like K$^{*}(892)^{0}$ and $\phi(1020)$ can also contribute to a 
systematic study of the enhancement of baryon-to-meson ratios (e.g., p/$\pi$ 
and $\Lambda/\mathrm{K}^{0}_{\mathrm{S}}$~\cite{Adler:2003kg, Abelev:2013xaa}) 
at intermediate $p_{\mathrm{T}}$. Recombination models suggest that the number 
of constituent quarks of the hadrons determine the enhancement, while 
hydrodynamic models explain this on the basis of differences in the hadron masses 
leading to different radial flow patterns. The K$^{*}(892)^{0}$ and $\phi(1020)$ 
mesons, which have masses very close to that of a proton, are well suited for 
testing the underlying hadron production mechanisms. 

In this paper, K$^{*}(892)^{0}$ and $\phi(1020)$ meson production in pp and 
Pb--Pb collisions at $\sqrt{s_\mathrm{NN}}$ = 2.76 TeV is studied. We have 
previously published measurements of K$^{*}(892)^{0}$ and $\phi(1020)$ 
meson production for $p_{\mathrm{T}}$ $\textless$ 5 GeV/$c$ in Pb--Pb 
collisions at $\sqrt{s_\mathrm{NN}}$ = 2.76 TeV~\cite{Abelev:2014uua} using 
data recorded in 2010. The high luminosity data taken by ALICE in 2011 allow 
statistically improved signal measurements. The spectra have been measured in 
the range 0 $\textless$ $p_{\mathrm{T}}$ $\textless$ 15 GeV/$c$ (0.4 $\textless$ 
$p_{\mathrm{T}}$ $\textless$ 21 GeV/$c$) in minimum bias pp collisions and 
0.3 $\textless$ $p_{\mathrm{T}}$ $\textless$ 20 GeV/$c$ (0.5 $\textless$ 
$p_{\mathrm{T}}$ $\textless$ 21 GeV/$c$) in Pb--Pb collisions in six (seven) 
centrality classes for K$^{*}(892)^{0}$ ($\phi(1020)$). This new data set 
also allowed the measurement of K$^{*}(892)^{0}$ in finer centrality intervals
in central and semi-central Pb--Pb collisions to study hadron production
mechanisms at low, intermediate and high $p_{\mathrm{T}}$. The new 
measurements of K$^{*}(892)^{0}$ and $\phi(1020)$ meson production in pp 
collisions at $\sqrt{s}$ = 2.76 TeV are used to calculate particle ratios 
and also to test various perturbative QCD inspired event generators.

The nuclear modification factor ($R_\mathrm{AA}$) is defined as the yield 
of particles in heavy-ion collisions relative to that in elementary pp 
collisions, scaled with the average nuclear overlap function.

\begin{equation}
  R_{\mathrm{AA}}=\frac{1}{\langle T_{\mathrm{AA}} \rangle} \times 
  \frac{(\mathrm{d}^{2}N/\mathrm{d}y \mathrm{d}p_{T})_{\mathrm{AA}}}
       {(\mathrm{d}^{2}\sigma/\mathrm{d}y \mathrm{d}p_{T})_{\mathrm{pp}}},
       \label{eq33}
\end{equation}

where $\langle T_{\mathrm{AA}} \rangle$ = $\langle N_{\mathrm{coll}} \rangle$
/ $\sigma_{\mathrm{inel}}$ is the average nuclear overlap function, 
$\langle N_{\mathrm{coll}} \rangle$ is the average number of binary 
nucleon-nucleon collisions calculated using MC Glauber~\cite{Miller:2007ri}
simulations and $\sigma_{\mathrm{inel}}$ is the inelastic pp cross section
~\cite{Abelev:2013qoq}.

Throughout this paper, the results for K$^{*}(892)^{0}$ and  
$\mathrm{\overline{K}}^{*}(892)^{0}$ are averaged and denoted by the 
symbol K$^{*0}$ and $\phi(1020)$ is denoted by $\phi$ unless specified 
otherwise. The paper is organized as follows: Section~\ref{sec:data} 
describes the data analysis techniques. Section~\ref{sec:results} 
presents results including K$^{*0}$ and $\phi$ meson $p_{\mathrm{T}}$ 
spectra, ratios to different hadrons and nuclear modification factors.
A summary is given in Section~\ref{sec:conc}. 

\section{Data analysis} \label{sec:data}
New measurements of K$^{*0}$ and $\phi$ meson production have been performed 
on data taken with the ALICE detector in the year 2011. The resonances are 
reconstructed via hadronic decay channels with large branching ratios (BR): 
K$^{*0}\rightarrow\pi^{\pm}\mathrm{K}^{\mp}$ with BR 66.6$\%$ and 
$\phi \rightarrow \mathrm{K}^{+}\mathrm{K}^{-}$ with BR 48.9$\%$~\cite{Agashe:2014kda}.
For both K$^{*0}$ and $\phi$, the measurements are performed in six common 
centrality classes, 0--5$\%$, 5--10$\%$, 10--20$\%$, 20--30$\%$, 30--40$\%$, 
40--50$\%$. The peripheral centrality class 60--80$\%$ is also measured for 
$\phi$ only. 

\subsection{Event and track selection} \label{sec:event}   
The data in pp collisions were collected in 2011 using a minimum bias (MB) 
trigger, requiring at least one hit in any of V0-A, V0-C, and Silicon Pixel 
Detectors (SPD), in coincidence with the presence of an LHC bunch 
crossing~\cite{Cortese:2004aa, Abbas:2013taa}. The ALICE V0 are small-angle 
plastic scintillator detectors placed on either side of the collision vertex, 
covering the pseudorapidity ranges 2.8 $<$ $\eta$ $<$ 5.1 (V0-A) and 
--3.7 $<$ $\eta$ $<$ --1.7 (V0-C). The two SPD layers , which cover $|\eta|$ 
$<$ 2.0, are the innermost part of the the Inner Tracking System (ITS), 
composed of six layers of silicon detector placed radially between 3.9 and 
43 cm around the beam pipe. During the high luminosity Pb--Pb run in 2011,
V0 online triggers are used to enhance central 0--10$\%$, semicentral 
10--50$\%$ and select MB (0--80$\%$) events. The trigger was 100$\%$ efficient 
for the 0--8$\%$ most central Pb--Pb collisions and 80$\%$ efficient for 
centrality 8--10$\%$~\cite{Adam:2016xbp}. The inefficiency for the 8--10$\%$ 
range has a negligible ($<$1$\%$) effect on the results presented in this
paper. The numbers of events after event selections is summarized in 
Table~\ref{tab_event}. 

\begin{table}
  \begin{center}
    \scalebox{1.0}{
      \begin{tabular}{ccccc}
        \hline \hline
        Centrality & Events & Year & Data Set\\
        \hline
        0--10$\%$  & 2.0 $\times$ $10^{7}$ & 2011 & Pb--Pb \\
        10--50$\%$ & 1.8 $\times$ $10^{7}$ & 2011 & Pb--Pb \\
        0--80$\%$  & 6.0 $\times$ $10^{5}$ & 2011 & Pb--Pb \\
        \hline
        MB         & 3.0 $\times$ $10^{7}$ & 2011 & pp     \\
        \hline \hline
      \end{tabular}
    }
  \end{center}
  \caption{Summary of different trigger selected data sets and number 
    of events analyzed in pp and Pb--Pb collisions at 
    $\sqrt{s_\mathrm{NN}}$ = 2.76 TeV.}
  \label{tab_event}
\end{table}

A detailed description of the ALICE detector is given in 
Refs.~\cite{Alessandro:2006yt, Aamodt:2008zz, Abelev:2014ffa}. The ALICE Inner 
Tracking System (ITS) and the Time Projection Chamber (TPC), are used for 
tracking and reconstruction of the primary vertex. Events are required to 
have the primary vertex coordinate along the beam axis ($v_{\mathrm{z}}$)
within 10 cm from the nominal interaction point. Tracks in the TPC are 
selected for both K$^{*0}$ and $\phi$ reconstruction with the requirement 
of at least 70 TPC pad rows measured along the track out of a maximum 
possible 159. The TPC covers the pseudorapidity range $|\eta|$ $\textless$ 
0.9 with full azimuthal acceptance. To ensure a uniform acceptance, the 
tracks are selected within $|\eta|$ $\textless$ 0.8. The data sample for 
the pp analysis is chosen to have minimal pileup; Pb--Pb collisions have
negligible pileup. In order to reduce contamination from beam-background 
events and secondary particles coming from weak decays, cuts on the 
distance of closest approach to the primary vertex in the $xy$ plane 
(DCA$_{xy}$) and $z$ direction (DCA$_z$) are applied. The value of 
DCA$_{xy}$ is required to be less than 7 times its resolution: 
(DCA$_{xy}$($p_{\mathrm{T}})$ $\textless$ 0.0105 + 
0.035$p_{\mathrm{T}}^{-1.1}$) cm ($p_{\mathrm{T}}$ in GeV/$c$) 
and DCA$_{z}$, is required to be less than 2 cm. The $p_{\mathrm{T}}$ of 
each track is restricted to be greater than 0.15 GeV/$c$ for K$^{*0}$ 
in pp and Pb--Pb collisions and for $\phi$ in pp collisions. For $\phi$ 
in Pb--Pb collisions the track $p_{\mathrm{T}}$ was required to be $>$ 0.75 
GeV/$c$ for the 0--5$\%$ centrality class and $>$ 0.5 GeV/$c$ otherwise. The 
higher $p_{\mathrm{T}}$ cut for the $\phi$ analysis without particle 
identification (PID) was needed to improve the signal-to-background ratio at 
low momentum. 

The TPC has been used to identify charged particles by measuring the 
specific ionization energy loss (d$E$/d$x$). For K$^{*0}$ reconstruction, 
both in pp and Pb--Pb collisions, pion and kaon candidates are required 
to have mean values of the specific energy loss in the TPC 
($\langle$d$E$/d$x\rangle$) within two standard deviations 
($2\sigma_{\mathrm{TPC}}$) of the expected d$E$/d$x$ values for each 
particle species over all momenta. In the case of $\phi$ meson 
reconstruction, two PID selection criteria depending on the 
$p_{\mathrm{T}}$ of the $\phi$ meson are used. In both pp and Pb--Pb 
collisions the narrow $\phi$ signal is extracted from the unidentified
two-particle invariant-mass distribution for $p_{\mathrm{T}}$ 
$\textgreater$ 1 GeV/$c$. In pp collisions the production of the $\phi$ 
meson is additionally measured with a $2\sigma_{\mathrm{TPC}}$ 
restriction on $\langle$d$E$/d$x\rangle$ for 0.4 $\textless$ 
$p_{\mathrm{T}}$ $\textless$ 5 GeV/$c$. The spectra measured without 
PID in Pb--Pb collisions are comparable with the published 2010 
results~\cite{Abelev:2014uua} obtained with PID. Measurements with 
and without PID are found to be in good agreement for both collision 
systems in the overlap region (1 $\textless$ $p_{\mathrm{T}}$ $\textless$ 
5 GeV/$c$). The $p_{\mathrm{T}}$ spectra in this paper are combinations
of results obtained with PID at low momentum ($p_{\mathrm{T}}$ $\textless$ 
3 GeV/$c$) and results obtained without PID for higher $p_{\mathrm{T}}$ 
in both pp and Pb--Pb collisions. 

\subsection{Yield extraction} \label{sec:signal}
The K$^{*0}$ ($\phi$) is reconstructed through its dominant hadronic decay 
channel by calculating the invariant-mass of its daughters at the primary 
vertex. The invariant-mass distribution of the daughter pairs is constructed 
using all unlike-sign pairs of charged K candidates with oppositely charged 
$\pi$ (K) candidates for K$^{*0}$ ($\phi$). The rapidity of $\pi$K and KK 
pairs is required to lie within the range $|y_{\mathrm{pair}}|$ $\textless$ 
0.5. 
                                                                        
\begin{figure}[H]
  \begin{center}
   \includegraphics[scale=0.54]{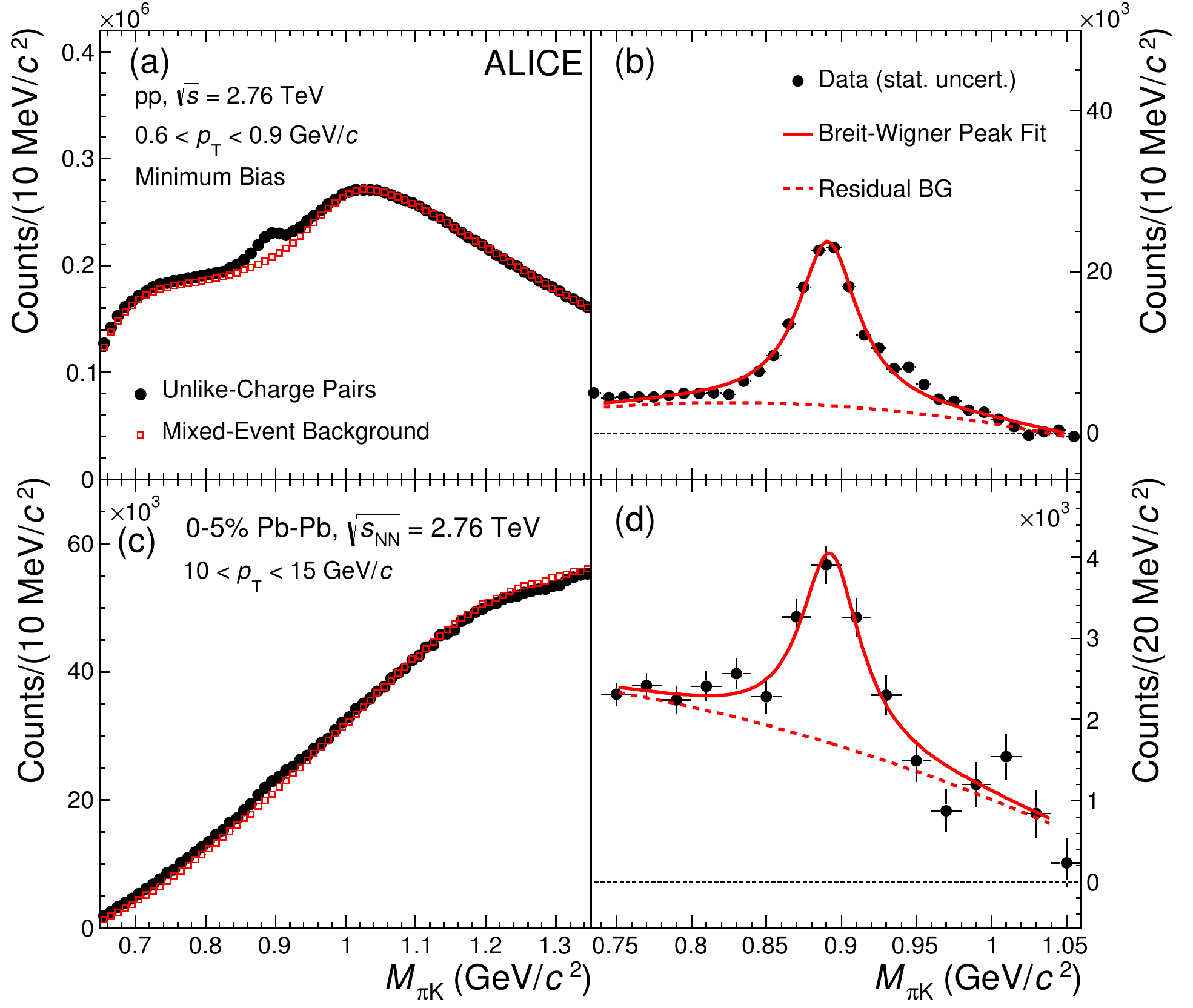}
    \caption{(Color online) Invariant-mass distributions of $\pi$K pairs for pp 
      and the 0--5$\%$ most central Pb--Pb collisions at $\sqrt{s_{\mathrm{NN}}}$ 
      = 2.76 TeV for the momentum ranges 0.6 $\textless$ $p_{\mathrm{T}}$ $\textless$ 
      0.9 GeV/$c$ (upper panel) and 10 $\textless$ $p_{\mathrm{T}}$ $\textless$ 
      15 GeV/$c$ (lower panel), respectively. Panels (a) and (c) show the unlike 
      charge $\pi$K invariant-mass distribution from the same event and normalized 
      mixed event background. Panels (b) and (d) report the invariant-mass 
      distribution after subtraction of the combinatorial background for 
      K$^{*0}$. The statistical uncertainties are shown by bars. The solid 
      curves represent fits to the distributions and the red dashed curves 
      are the components of those fits that describe the residual background.
    }
    \label{signal_kstar}
  \end{center}
\end{figure}

The signal extraction follows the procedure of the already published 
analysis~\cite{Abelev:2014uua}. The combinatorial background is estimated 
using the event mixing technique by pairing decay daughter candidates from 
two different events with similar primary vertex positions ($v_\mathrm{z}$) 
and centrality percentiles in Pb--Pb collisions. For the K$^{*0}$ analysis, 
the difference in the event plane angles between two events is required to 
be less than 30$^{\circ}$. The Pb--Pb data sample is divided into 10 bins in 
centrality percentiles and 20 bins in $v_\mathrm{z}$. Each event is mixed 
with 5 other similar events for both $\pi$K and KK. For event mixing in pp 
collisions, the binning takes into account the multiplicity of charged 
particles measured using the TPC. The total multiplicity and $v_\mathrm{z}$ 
are divided in 10 bins each for both $\pi$K and KK. These requirements ensure 
that the mixed events have similar features, so the invariant-mass distribution 
from the event mixing can better reproduce the combinatorial background.

\begin{figure}[H]
  \begin{center}                      
    \includegraphics[scale=0.54]{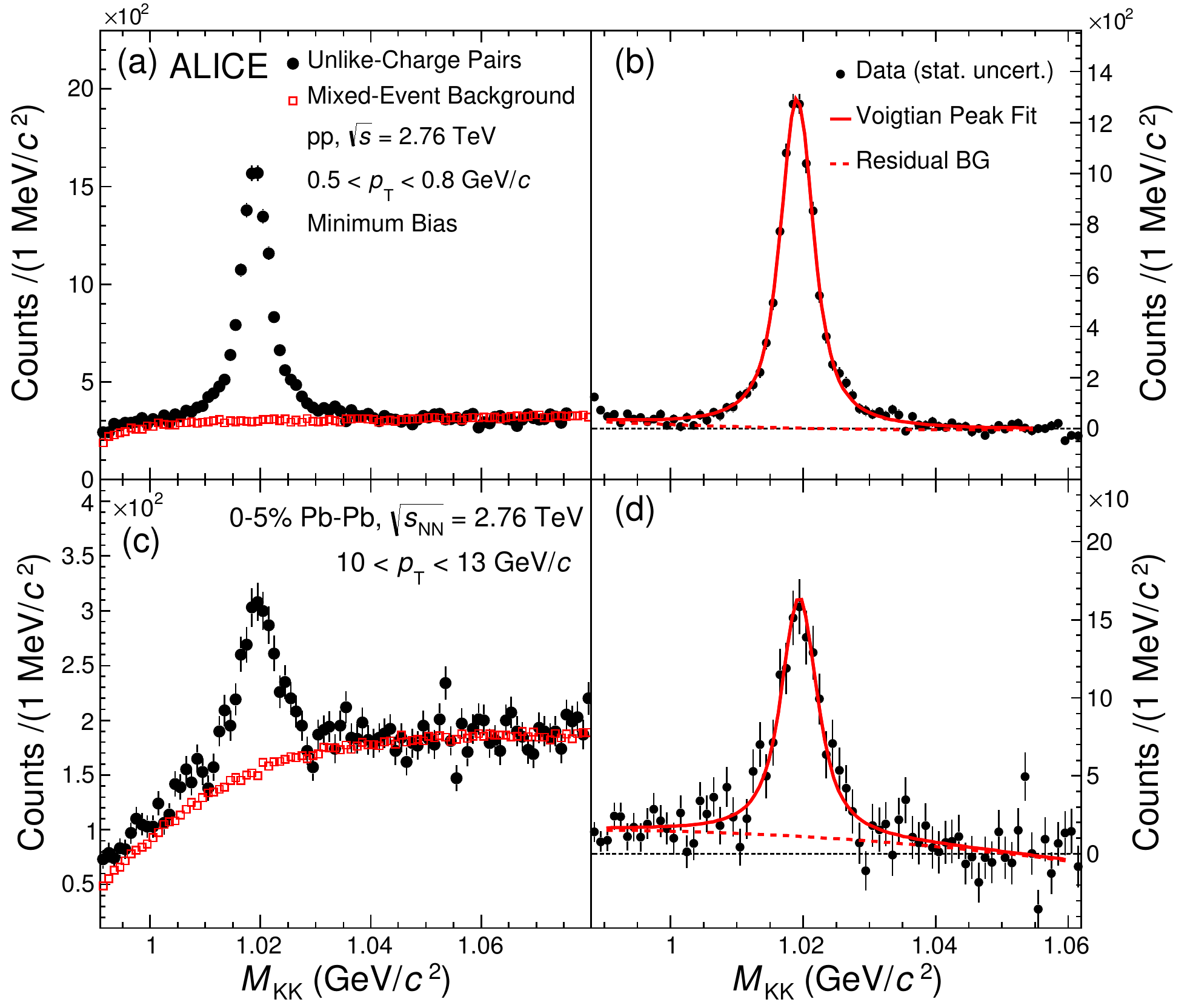}
    \caption{(Color online) Invariant-mass distributions of KK pairs for pp and 
      the 0--5$\%$ most central Pb--Pb collisions at $\sqrt{s_{\mathrm{NN}}}$ = 
      2.76 TeV for the momentum ranges 0.5 $\textless$ $p_{\mathrm{T}}$ $\textless$ 
      0.8 GeV/$c$ (upper panel) and 10 $\textless$ $p_{\mathrm{T}}$ $\textless$ 
      13 GeV/$c$ (lower panel), respectively. In panels (a) and (c) the unlike 
      charge KK invariant-mass distribution from the same event and normalized
      mixed event background are shown. In panels (b) and (d) the invariant-mass 
      distribution after subtraction of the combinatorial background for $\phi$
      is shown. The statistical uncertainties are shown by bars. The solid 
      curves are the fits to the distributions and the red dashed curves are 
      the components of those fits that describe the residual background.}
    \label{signal_phi}
  \end{center}
\end{figure}

In Fig.~\ref{signal_kstar} (Fig.~\ref{signal_phi}), panels (a) and (c)
show the $\pi^{\mp}$K$^{\pm}$ (K$^{+}$K$^{-}$) invariant-mass distributions 
from the same event and mixed events for 0.6 $\textless$ $p_{\mathrm{T}}$
$\textless$ 0.9 GeV/$c$ (0.5 $\textless$ $p_{\mathrm{T}}$ $\textless$ 
0.8 GeV/$c$) in minimum bias pp collisions and 10 $\textless$ $p_{\mathrm{T}}$ 
$\textless$ 15 GeV/$c$ (10 $\textless$ $p_{\mathrm{T}}$ $\textless$ 13 GeV/$c$) 
in 0--5$\%$ central Pb--Pb collisions at $\sqrt{s_\mathrm{NN}}$ = 2.76 TeV. 
The mixed event distribution is normalized to the same event distribution in 
the invariant-mass region of 1.1 to 1.3 GeV/$c^{2}$ (1.04 to 1.06 GeV/$c^{2}$),
which is away from the signal peaks. The $\pi^{\mp}$K$^{\pm}$ (K$^{+}$K$^{-}$) 
invariant-mass distributions after mixed event background subtraction are shown 
in panels (b) and (d) of Fig.~\ref{signal_kstar} (Fig.~\ref{signal_phi}), 
where the signals are observed on top of a residual background. The residual 
background is due to correlated $\pi$K or KK pairs emitted within jets and 
from mis-reconstructed hadronic decays~\cite{Abelev:2014uua}. The shape of 
the residual background is studied by means of Monte Carlo simulations. It 
exhibits a smooth dependence on mass and a second order polynomial is found 
to be a suitable function to describe the residual background for both K$^{*0}$ 
and $\phi$.

For each $p_{\mathrm{T}}$ interval and collision centrality class, the 
invariant-mass distribution is fitted with the sum of a peak fit function and 
a second-order polynomial to account for the residual background. The $\pi$K 
distribution signal peak is parametrized with a Breit-Wigner function. The fit 
function for K$^{*0}$ is

\begin{equation}
  \frac{\mathrm{d}N}{\mathrm{d}m_{\pi \mathrm{K}}} = \frac{Y}{2\pi} \times 
  \frac{\Gamma_0}{(m_{\pi \mathrm{K}} - M_0)^{2} + \frac{\Gamma_{0}^{2}}{4}} 
  + (Am_{\pi \mathrm{K}}^{2} + Bm_{\pi \mathrm{K}} + C), \label{eq1}
\end{equation}

where $M_{\mathrm{0}}$ is the reconstructed mass of K$^{*0}$, $\Gamma_{\mathrm{0}}$ 
is the resonance width fixed to the value in vacuum~\cite{Agashe:2014kda} and $Y$ is 
yield of the K$^{*0}$ meson. The mass resolution of the K$^{*0}$ is 
negligible compared to its width (47.4 $\pm$ 0.6 MeV/$c^{2}$) and is therefore 
not included in the K$^{*0}$ fitting function. $A$, $B$ and $C$ are the 
polynomial fit parameters. Similarly, for the KK signal peak is fitted with
a Voigtian function (a Breit-Wigner function convoluted with a Gaussian function) 
is used, which accounts for the resonance width and the detector mass resolution.
The fit function for $\phi$ is

\begin{equation}
  \frac{\mathrm{d}N}{\mathrm{d}m_{\mathrm{KK}}} = \frac{Y \Gamma_{0}}
       {(2\pi^{3/2})\sigma} \times \int_{-\infty}^{+\infty} \mathrm{exp} 
       \Bigg(\frac{(m_{\mathrm{KK}} - m')^{2}}{2\sigma^{2}}\Bigg) 
       \frac{1}{(m'-M_{0})^{2} + \frac{\Gamma_{0}^{2}}{4} } \mathrm{d}m' 
       + (Am_{\mathrm{KK}}^{2} + Bm_{\mathrm{KK}} + C),
       \label{eq2}
\end{equation}

where the parameter $\sigma$ is the $p_{\mathrm{T}}$-dependent mass resolution, 
which is found to be independent of collision centrality. For Pb--Pb (pp) 
collisions, the mass resolution parameter has been extracted by using HIJING 
(PYTHIA)~\cite{Wang:1991hta, Sjostrand:2006za} simulations, where the decay 
products of $\phi$ are propagated through the ALICE detector, by using 
GEANT3~\cite{Brun:1994aa}.

The $\pi^{\mp}$K$^{\pm}$ (K$^{+}$K$^{-}$) invariant-mass distribution is fitted 
in the range 0.75 $\textless$ $m_{\pi \mathrm{K}}$ $\textless$ 1.05 GeV/$c^{2}$ 
(0.99 $\textless$ $m_{\mathrm{KK}}$ $\textless$ 1.06 GeV/$c^{2}$). The yield of 
K$^{*0}$ ($\phi$) is extracted in each $p_{\mathrm{T}}$ interval and centrality 
class by integrating the mixed-event background subtracted invariant-mass
distribution in the range 0.77 $\textless$ $m_{\pi\mathrm{K}}$ $\textless$ 
1.02 GeV/$c^{2}$ (1 $\textless$ $m_{\mathrm{KK}}$ $\textless$ 1.03 GeV/$c^{2}$), 
subtracting the integral of the residual background function in the same range 
and correcting the result to account for the yields outside this range. This 
correction to the total yield is about 9$\%$ (13$\%$) for K$^{*0}$ 
($\phi$)~\cite{Abelev:2014uua}. 

\subsection{Yield correction} \label{sec:yield}
The raw yields of K$^{*0}$ and $\phi$ mesons are normalized to the number of events 
and corrected for the branching ratio (BR)~\cite{Agashe:2014kda}, the detector 
acceptance ($A$) and the reconstruction efficiency ($\epsilon_{\mathrm{rec}}$). 

\subsubsection{Acceptance and reconstruction efficiency} \label{sec:eff}
\begin{figure}[ht]
  \begin{center}
   \includegraphics[scale=0.35]{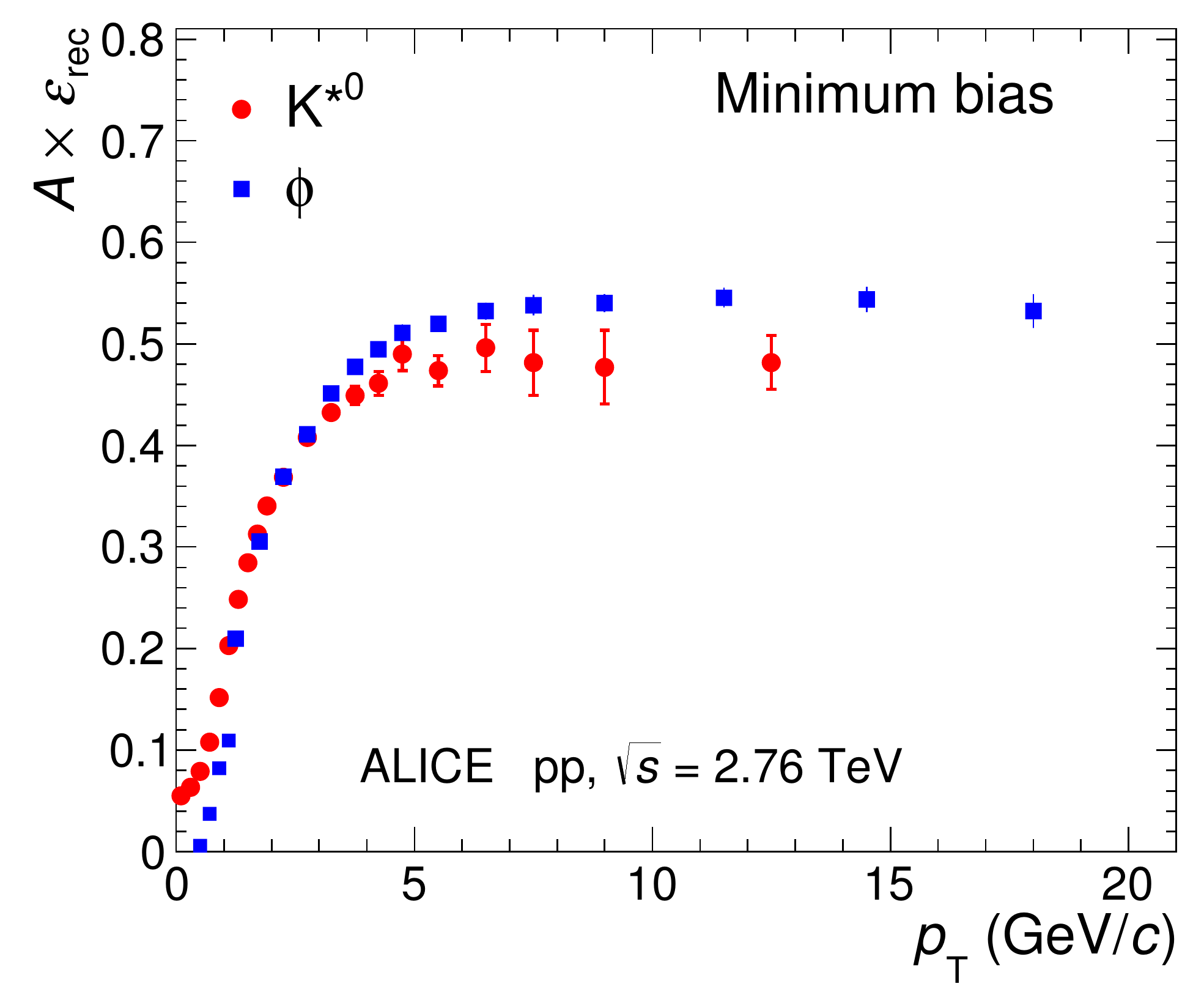}
   \includegraphics[scale=0.35]{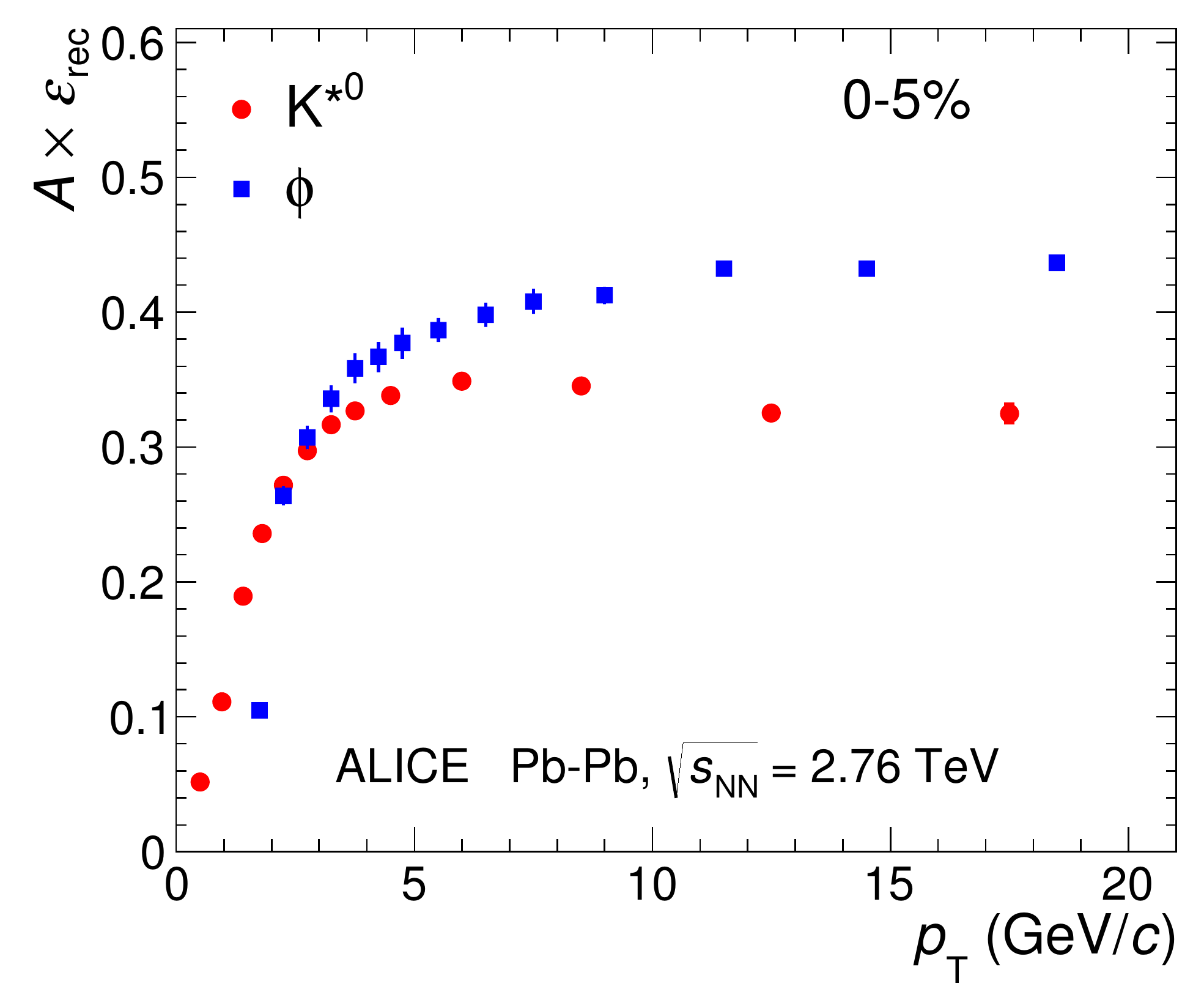}
    \caption{(Color online) The acceptance and efficiency 
      ($A \times$ $\epsilon_{\mathrm{rec}}$) correction as a function of 
      $p_{\mathrm{T}}$ for K$^{*0}$ (red marker) and $\phi$ (blue marker) 
      mesons in pp (left panel) and 0-5$\%$ centrality in Pb--Pb (right panel) 
      collisions at $\sqrt{s_\mathrm{NN}}$ = 2.76 TeV.}
    \label{eff_pp_pbpb}
  \end{center}
\end{figure}

A Monte Carlo simulation based on the HIJING (PYTHIA) event generator is 
used for the estimation of the acceptance $\times$ efficiency ($A \times$ 
$\epsilon_{\mathrm{rec}}$) in Pb--Pb (pp) collisions. Figure~\ref{eff_pp_pbpb} 
shows $A \times$ $\epsilon_{\mathrm{rec}}$ for minimum bias pp collisions and 
0-5$\%$ centrality Pb--Pb collisions at $\sqrt{s_\mathrm{NN}}$ = 2.76 TeV for 
both K$^{*0}$ and $\phi$. In these simulations, the decay products of the 
generated K$^{*0}$ and $\phi$ are propagated through the ALICE detector material
using GEANT3~\cite{Brun:1994aa}. The $A \times$ $\epsilon_{\mathrm{rec}}$ is 
defined as the fraction of generated K$^{*0}$ and $\phi$ that is reconstructed
after passing through the detector simulation, the event reconstruction and 
being subjected to the track quality, PID and pair rapidity cuts. In this 
calculation, only those K$^{*0}$ ($\phi$) mesons that decay to K$^{\pm}\pi^{\mp}$ 
(K$^{+}$K$^{-}$) are used. The correction for the branching ratio is therefore 
not included in $A \times$ $\epsilon_{\mathrm{rec}}$ and is applied separately 
(Eq.~\ref{eq3}). The differences in $A \times$ $\epsilon_{\mathrm{rec}}$ for 
K$^{*0}$ and $\phi$ are due to the different kinematics and track selection 
criteria. In Pb--Pb collisions, $A \times$ $\epsilon_{\mathrm{rec}}$ has a 
very mild centrality dependence. 

\subsubsection{Normalization} \label{sec:norm}
The yields are normalized to the number of minimum bias events and corrected 
for the trigger ($\epsilon_{\mathrm{trigger}}$) and vertex reconstruction 
efficiencies ($\epsilon_{\mathrm{vertex}}$) to obtain the absolute resonance 
yields per inelastic pp collision. The $\epsilon_{vertex}$ correction was 
estimated to be equal to 89$\%$ and takes into account K$^{*0}$ and $\phi$ 
meson losses after imposing the vertex cut. The trigger efficiency correction 
factor $\epsilon_{\mathrm{trigger}}$ is 88.1$\%$ with relative uncertainty of
+5.9$\%$ and -3.5$\%$ for pp collisions~\cite{Abelev:2012sea}. The effects of
trigger and vertex reconstruction efficiency corrections are negligible in 
Pb--Pb collisions and, hence, not considered. The invariant yield for pp and
 Pb--Pb collisions is 

\begin{equation}
  \frac{1}{2\pi p_{\mathrm{T}}}\frac{\mathrm{d}^{2}N}{\mathrm{d}y\mathrm{d}
    p_{\mathrm{T}}} =\frac{1}{2 \pi p_{\mathrm{T}}} \times \frac{1}
       {N_{\mathrm{ev}}} \times \frac{N^{\mathrm{raw}}}{\mathrm{d}y\mathrm{d}
         p_{\mathrm{T}}} \times \frac{\epsilon_{\mathrm{trigger}}}{A \times 
         \epsilon_{\mathrm{rec}} \times BR \times \epsilon_{\mathrm{vertex}}}
       \label{eq3}
\end{equation}

where $N_{\mathrm{ev}}$ is the number of events used in the analysis and 
{$\it{N}^{raw}$} is the K$^{*0}$ or $\phi$ raw yield. 

\subsection{Systematic uncertainties} \label{sec:syst}
The sources of systematic uncertainties in the measurement of K$^{*0}$ 
and $\phi$ production in pp and Pb--Pb collisions are the global tracking
(performed using ITS and TPC clusters) efficiency, track selection cuts,
PID, yield extraction method and material budget. In Pb--Pb (pp) collisions, 
the uncertainty contribution due to the global tracking efficiency has been 
estimated to be 5$\%$ (4$\%$) for charged particles~\cite{Abelev:2014laa},
which results in a 10$\%$ (8$\%$) effect for the track pairs used for the 
invariant-mass analysis of K$^{*0}$ and $\phi$. The systematic uncertainty 
in the global tracking efficiency of the charged decay daughters is 
$p_{\mathrm{T}}$ and centrality independent and it cancels out partially 
in particle yield ratios for both K$^{*0}$ and $\phi$. The uncertainty due 
to the PID cuts is 3.7$\%$ (4$\%$) in pp and 4$\%$ (6.2$\%$) in Pb--Pb 
collisions for K$^{*0}$ ($\phi$). Systematic uncertainties of 3$\%$ to 
6$\%$ on the raw yield have been assigned due to variation of the track 
selection cuts, depending on the particle species and collision system. 
The uncertainty due to the raw yield extraction includes variations of the 
fit range, fit function, mass resolution and mixed event background 
normalization range. The $\pi$K (KK) invariant-mass fitting ranges were 
varied by 10--30 (5--10) MeV/$c^{2}$ on each side of the peak. The residual
background is fitted with a 3$^{rd}$-order polynomial and the resulting 
variations in the raw yield are also incorporated into the systematic
uncertainties. Due to the uncertainty in the material budget of the ALICE
detectors, a systematic uncertainty of $\sim$1$\%$ (derived from the study 
for $\pi^{\pm}$ and K$^{\pm}$ in~\cite{Abelev:2014laa}) is added to the yield
of K$^{*0}$ and $\phi$ at low $p_{\mathrm{T}}$ $\textless$ 2 GeV/$c$, the
contribution is negligible at higher $p_{\mathrm{T}}$. For $\phi$ the change 
in the yield due to a variation of the mass resolution is included in the 
systematic uncertainties of the raw yield extraction. The systematic
uncertainties due to yield extraction are 2.5--14$\%$ (2--13$\%$) for 
K$^{*0}$ ($\phi$) in pp collisions and 4--15$\%$ (3.5--13$\%$) for K$^{*0}$
($\phi$) in Pb--Pb collisions. Raw yield extraction dominates total 
uncertainties in the lowest and highest $p_{\mathrm{T}}$ intervals. All 
other systematic uncertainties have weak $p_{\mathrm{T}}$ and centrality 
dependence, with the exception of the yield extraction uncertainty. The 
total systematic uncertainties amount to 10--18$\%$ (9--16$\%$) 
for K$^{*0}$ ($\phi$) in pp collisions and 12--19$\%$ (13--18$\%$) for 
K$^{*0}$ ($\phi$) in Pb--Pb collisions. The contributions are summarized 
in Table~\ref{tab_systematic}. 
 
    \begin{table}
      \begin{center}
        \scalebox{0.9}{
          \begin{tabular}{cccccc}
            \hline \hline
            & ~~~~~~~~~~~~~~~~~~~~~~~~~ Pb--Pb    &    &    & ~~~~~~~~~~~~~~~~ pp   &        \\
            \hline 
            Systematic variation        &  K$^{*0}$  &  $\phi$    &    &  K$^{*0}$  &  $\phi$\\
            \hline 
            Global tracking efficiency  &    10      &     10     &    &   8        &   8    \\
            Track selection             &    3-6     &    3-5     &    &   3        &   3    \\
            Particle identification     &    4.0     &    6.2     &    &   3.7      &  1-4   \\
            Material budget             &$\textless$1&$\textless$1&    &  0-3.3     &  0-3.3 \\
            Yield extraction            &    4-15    &   3.5-13   &    &  2.5-14    &  2-13  \\
            \hline
            Total                       &   12-19    &   13-18    &    &  10-18     &  9-16   \\
            \hline \hline
          \end{tabular}
        }
      \end{center}
      \caption{Systematic uncertainties in the measurement of K$^{*0}$ and $\phi$ 
        yields in pp and Pb--Pb collisions at $\sqrt{s_\mathrm{NN}}$ = 2.76 TeV. 
        The global tracking uncertainty is $p_{\mathrm{T}}$-independent, while 
        the other single valued systematic uncertainties are averaged over 
        $p_{\mathrm{T}}$. The values given in ranges are minimum and maximum 
        uncertainties depending on $p_{\mathrm{T}}$ and centrality class.
        The normalization uncertainty, which is due to uncertainties in the 
        boundaries of the centrality percentiles, are taken 
        from~\cite{Abelev:2013vea}.
      }
      \label{tab_systematic}
    \end{table}
    
    \section{Results}\label{sec:results}
    \subsection{$p_{\mathrm{T}}$ spectra in pp collisions} \label{sec:ppcross}
\begin{figure}
      \begin{center}
        \includegraphics[scale=0.38]{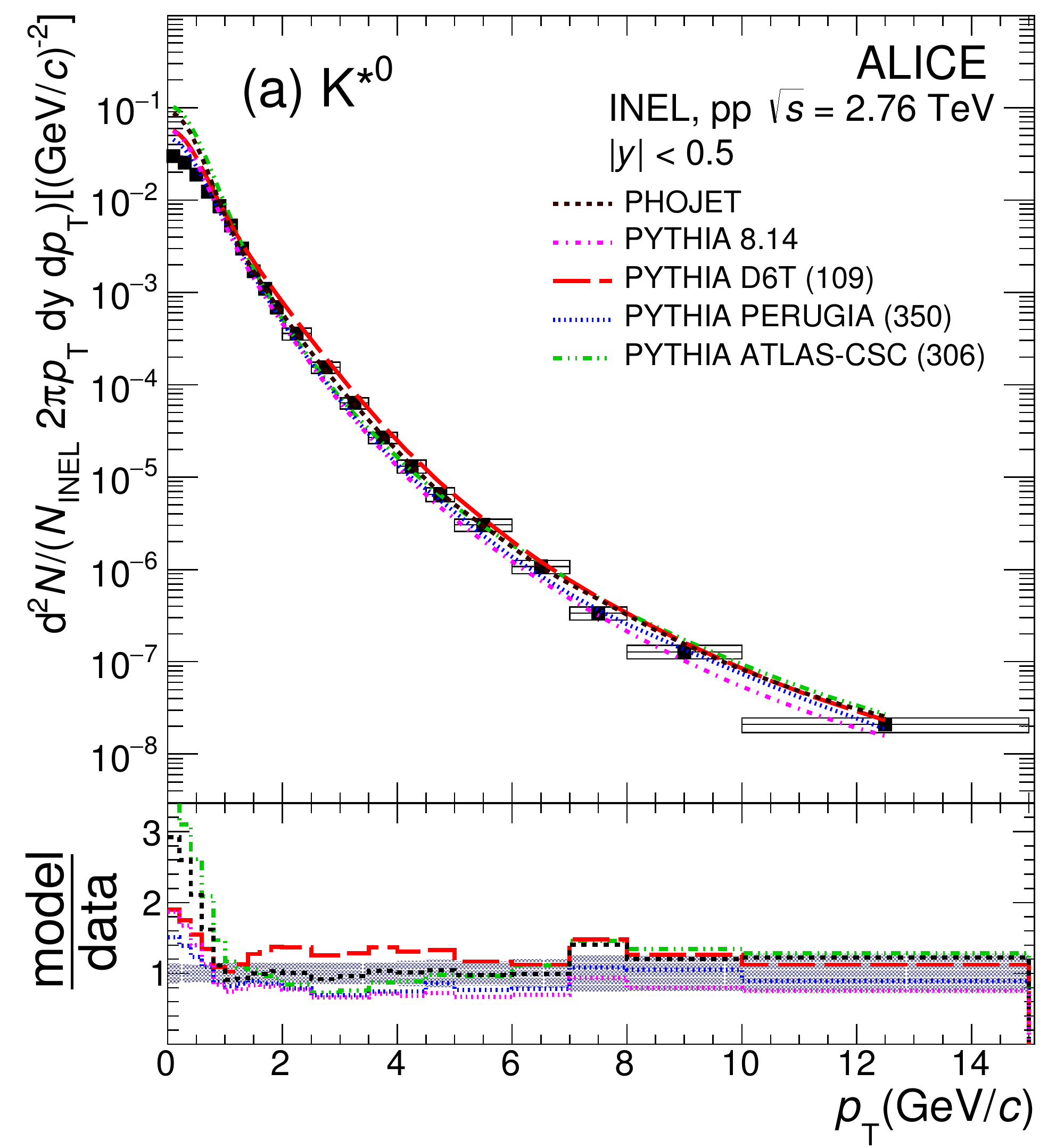}
        \includegraphics[scale=0.38]{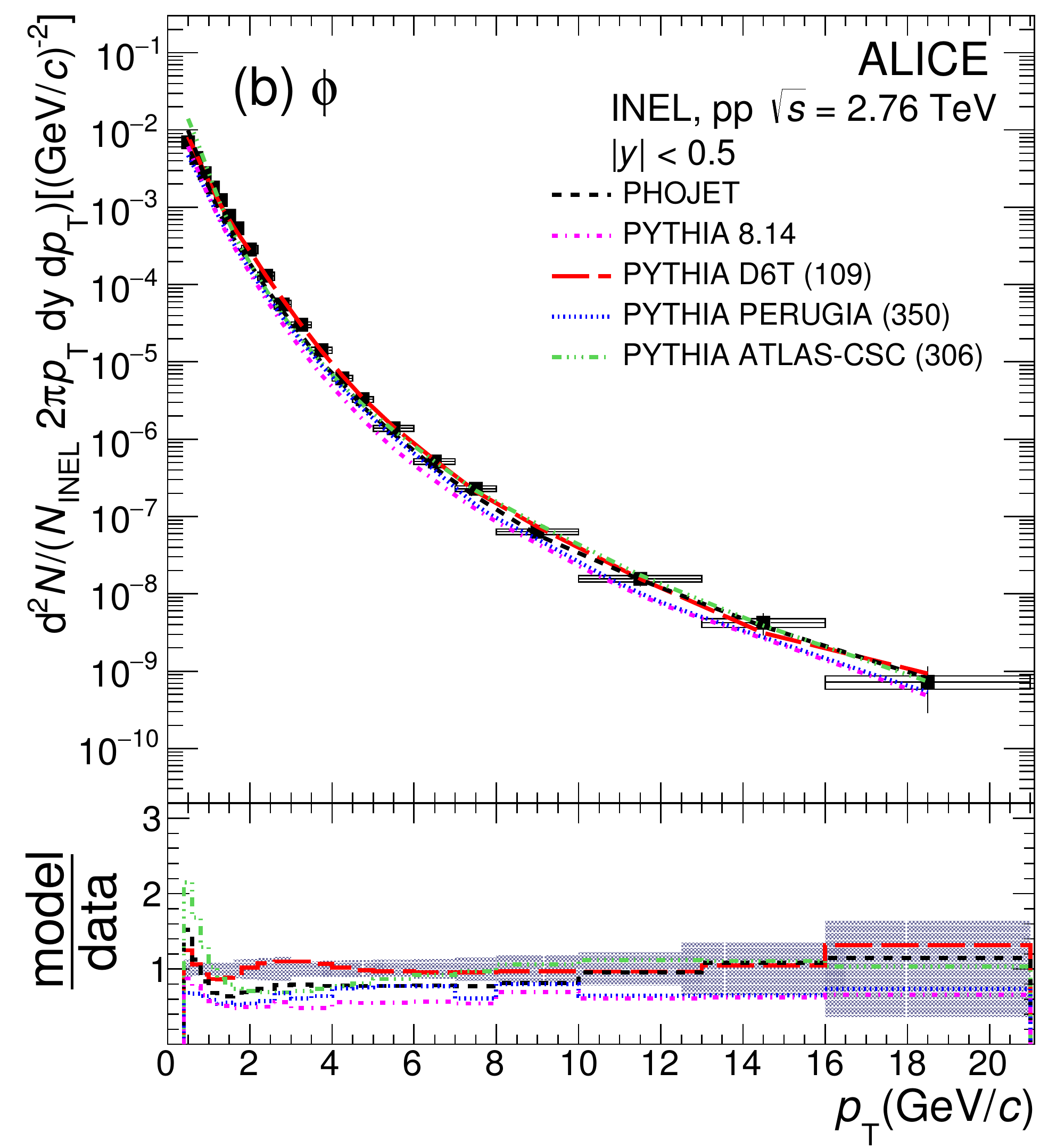}
        \caption{(Color online) Invariant yields for (a) K$^{*0}$ and (b) $\phi$ 
          mesons normalized to the number of inelastic pp collisions at
          $\sqrt{s}$ = 2.76 TeV. Invariant yield is calculated by taking the  
          value of $p_{\mathrm{T}}$ at the corresponding bin center. The 
          statistical uncertainties on the data are shown by bars and the 
          systematic uncertainties by boxes. The results are compared with model
          calculations from PYTHIA 8.14~\cite{Sjostrand:2007gs}, 
          PHOJET~\cite{Engel:1994vs, Engel:1995yda}, PYTHIA D6T~\cite{Field:2008zz},
          PYTHIA ATLAS-CSC~\cite{Buttar:2004iy} and PYTHIA PERUGIA~\cite{Skands:2010ak}
          as shown by different dashed lines. The lower panel for both K$^{*0}$ 
          and $\phi$ shows the model to data ratio.}
        \label{pp_pythia_phojet}
      \end{center}
    \end{figure}

    The first measurement of K$^{*0}$ ($\phi$) meson production in pp collisions 
    at $\sqrt{s}$ = 2.76 TeV up to $p_{\mathrm{T}}$ = 15 (21) GeV/$c$ is reported 
    here. Figure~\ref{pp_pythia_phojet} shows the transverse momentum spectra
    of K$^{*0}$ and $\phi$ mesons in pp collisions at $\sqrt{s}$ = 2.76 TeV, which 
    are compared with the values given by perturbative QCD inspired Monte Carlo 
    event generators PYTHIA~\cite{Sjostrand:2006za, Sjostrand:2007gs} and 
    PHOJET~\cite{Engel:1994vs, Engel:1995yda}. In both event generators hadronization 
    is simulated using the Lund String fragmentation model~\cite{Andersson:1983ia}. 
    Different PYTHIA tunes were developed by different groups through extensive comparison 
    of Monte Carlo distributions with the minimum bias data from various experiments. 
    The PYTHIA D6T tune~\cite{Field:2008zz} is adjusted to CDF Run 2 data, whereas the 
    ATLAS-CSC tune~\cite{Buttar:2004iy} is adjusted using UA5, E375 and CDF data from 
    $\sqrt{s}$ = 0.2 to 1.8 TeV. The Perugia tune~\cite{Skands:2010ak} uses the minimum 
    bias and underlying event data from the LHC at 0.9 and 7 TeV. The bottom panels
    in Fig.~\ref{pp_pythia_phojet} shows the ratio of the model calculations to the
    data. For the K$^{*0}$ meson, at low $p_{\mathrm{T}}$ ($\textless$ 1 GeV/$c$): 
    all models overpredict the data. In the intermediate $p_{\mathrm{T}}$ range 
    ($\sim$2--8 GeV/$c$): the Perugia, ATLAS-CSC and PYTHIA 8.14 tunes underestimate 
    the data, the D6T tune overestimates the data while PHOJET has good 
    agreement with the data. For the $\phi$ meson, at low $p_{\mathrm{T}}$ 
    ($\textless$ 1 GeV/$c$): PHOJET and ATLAS-CSC tune overpredict; the Perugia tune  
    and PYTHIA 8.14 underpredict the data. In the intermediate $p_{\mathrm{T}}$ 
    range ($\sim$2--8 GeV/$c$): the Perugia tune, PYTHIA 8.14, and PHOJET 
    underestimate the data, while the D6T and ATLAS-CSC tunes are in good agreement 
    with the data. In the high $p_{\mathrm{T}}$ range ($\textgreater$ 8 GeV/$c$)
    all models agree with the data within the uncertainties for both K$^{*0}$ and 
    $\phi$. For both K$^{*0}$ and $\phi$ mesons, the deviations of these models 
    from ALICE measurements are similar at both $\sqrt{s}$ = 2.76 and 
    7 TeV~\cite{Abelev:2012hy}.

    \subsection{$p_{\mathrm{T}}$ spectra in Pb--Pb collisions} \label{sec:pTspectra}
\begin{figure}
      \begin{center}
        \includegraphics[scale=0.39]{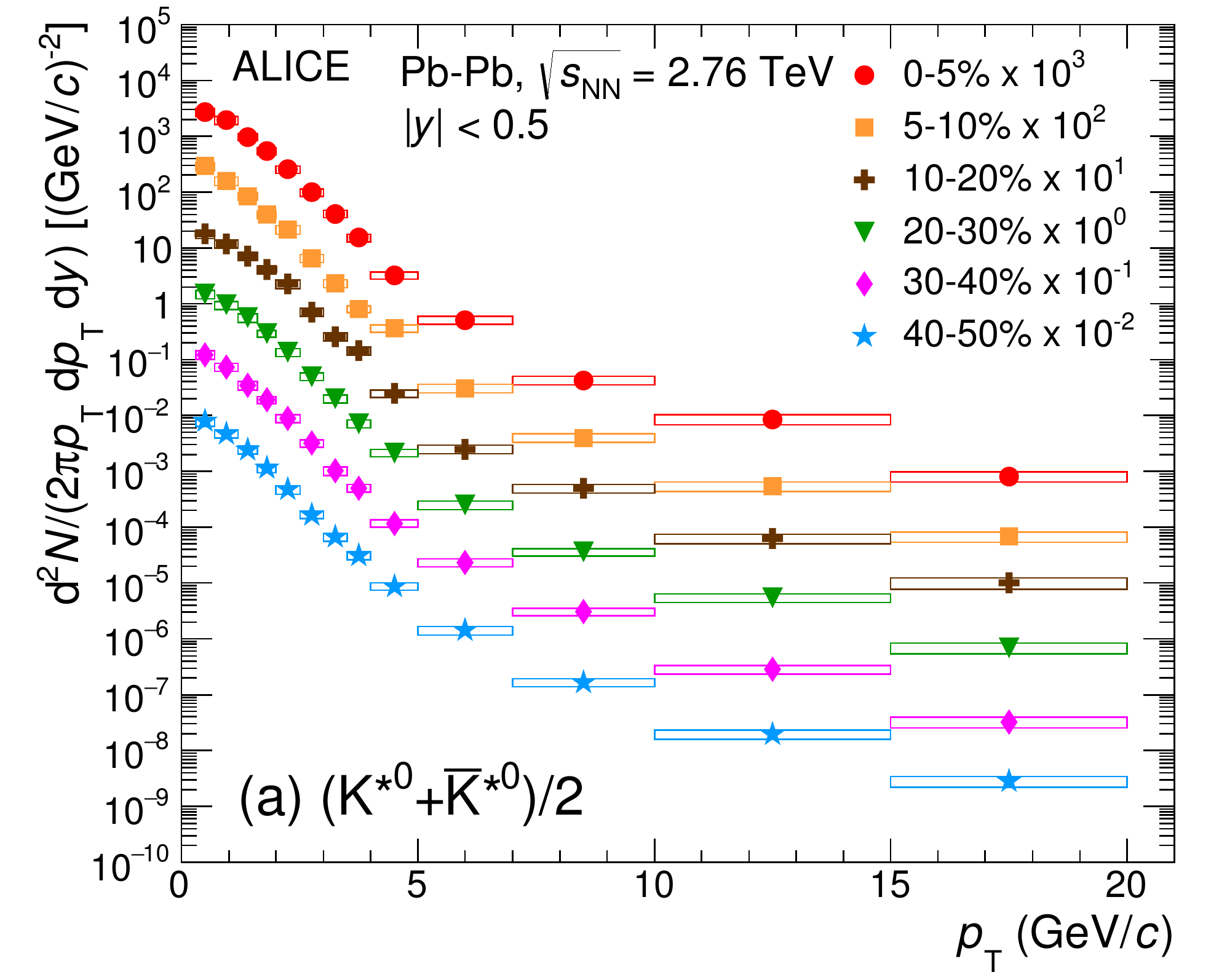}
        \includegraphics[scale=0.39]{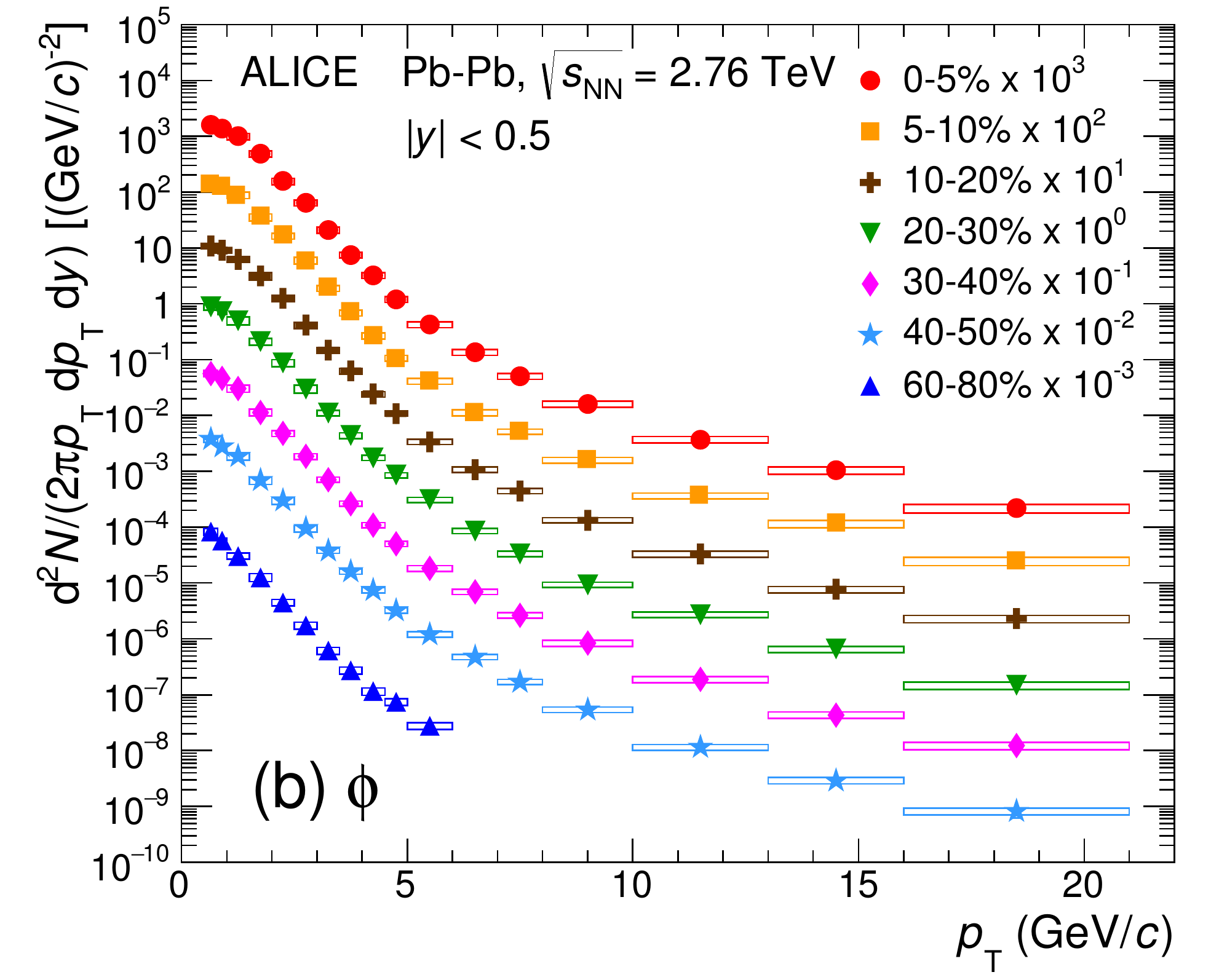}
        \caption{(Color online) Invariant yields of (a) K$^{*0}$ and (b) $\phi$ mesons in 
          various centrality classes in Pb--Pb collisions at $\sqrt{s_\mathrm{NN}}$ = 2.76 TeV. 
          Invariant yield is calculated by taking the value of $p_{\mathrm{T}}$ at the 
          corresponding bin center. The statistical and systematic uncertainties are 
          shown as bars and boxes, respectively. The normalization uncertainty is not 
          shown here, but is given in Table~\ref{tab_kstar_by_k}.}
        \label{pbpb_spectrA}
      \end{center}
    \end{figure}

    Figure~\ref{pbpb_spectrA} shows the $p_{\mathrm{T}}$ spectra for K$^{*0}$ and 
    $\phi$ mesons for different centrality classes in Pb--Pb collisions at 
    $\sqrt{s_\mathrm{NN}}$ = 2.76 TeV. The new measurements extend the previous 
    results~\cite{Abelev:2014uua} from $p_{\mathrm{T}}$ = 5 GeV/$c$ to 20 (21) GeV/$c$ 
    for K$^{*0}$ ($\phi$). The production of K$^{*0}$ has been measured in finer 
    centrality bins and compared to previously published results~\cite{Abelev:2014uua}. 
    When centrality bins are combined, the 2011 results are consistent with the 2010 data.    
    
    \subsection{Particle ratios}\label{sec:paricle}
    The measurements of K$^{*0}$ and $\phi$ spectra over a wide $p_{\mathrm{T}}$ 
    range are used to probe particle production mechanisms at different 
    $p_{\mathrm{T}}$ scales. The $p_{\mathrm{T}}$-integrated particle yield 
    (d$N$/d$y$) and the mean transverse momentum ($\langle{p_{\mathrm{T}}}\rangle$) 
    have been extracted using the procedure described in Ref.~\cite{Abelev:2014uua}. 
    The $p_{\mathrm{T}}$ distributions are fitted with a L\'evy-Tsallis 
    function~\cite{Tsallis:1987eu, Abelev:2006cs} in pp and a Boltzmann-Gibbs 
    blast-wave function~\cite{Schnedermann:1993ws} in Pb--Pb collisions. 
    The d$N$/d$y$ and $\langle{p_{\mathrm{T}}}\rangle$ have been extracted from the 
    data in the measured $p_{\mathrm{T}}$ region and the fit functions have been
    used to extrapolate into the unmeasured (low $p_{\mathrm{T}}$) region. 
    The low-$p_{\mathrm{T}}$ extrapolation covers $p_{\mathrm{T}}$ $<$ 0.3 (0.5) 
    GeV/$c$ for K$^{*0}$ ($\phi$) and accounts for 5$\%$ (14$\%$) of the total 
    yield. The yield is negligible at high-$p_{\mathrm{T}}$ ($>$ 20 GeV/$c$). These 
    values for K$^{*0}$ in pp and Pb--Pb collisions and the values for $\phi$ in pp 
    collisions are listed in Table~\ref{tab_kstar_by_k}. \\
  
    \begin{table}[ht]
      \scalebox{0.92}{
        \begin{tabular}{ccccc}
          \hline    \hline
          &  \multicolumn{2}{c}{K$^{*0}$ (Pb--Pb $\sqrt{s_\mathrm{NN}}$ = 2.76 TeV}\\
          \hline 
          Centrality ($\%$) & d$N$/d$y$  & K$^{*0}$/K$^{-}$  & $\langle{p_{\mathrm{T}}}\rangle$ (GeV/$c$)\\
          \hline
          0--5    &  19.56 $\pm$ 0.93 $\pm$ 2.48 $\pm$ 0.097  &  0.180 $\pm$ 0.008 $\pm$ 0.026 (0.023)  &  1.310 $\pm$ 0.023 $\pm$ 0.055\\
          5--10   &  16.71 $\pm$ 0.65 $\pm$ 2.08 $\pm$ 0.083  &  0.186 $\pm$ 0.007 $\pm$ 0.026 (0.024)  &  1.252 $\pm$ 0.023 $\pm$ 0.055\\
          10--20  &  13.65 $\pm$ 0.63 $\pm$ 1.84 $\pm$ 0.009  &  0.200 $\pm$ 0.009 $\pm$ 0.026 (0.023)  &  1.360 $\pm$ 0.026 $\pm$ 0.053\\
          20--30  &  10.37 $\pm$ 0.50 $\pm$ 1.38 $\pm$ 0.010  &  0.225 $\pm$ 0.011 $\pm$ 0.025 (0.023)  &  1.322 $\pm$ 0.028 $\pm$ 0.053\\
          30--40  &  7.35  $\pm$ 0.28 $\pm$ 0.97 $\pm$ 0.146  &  0.245 $\pm$ 0.009 $\pm$ 0.025 (0.021)  &  1.254 $\pm$ 0.023 $\pm$ 0.050\\
          40--50  &  4.66  $\pm$ 0.20 $\pm$ 0.65 $\pm$ 0.111  &  0.258 $\pm$ 0.011 $\pm$ 0.025 (0.022)  &  1.220 $\pm$ 0.025 $\pm$ 0.050\\
          \hline \\ 
          &  \multicolumn{2}{c}{K$^{*0}$ (pp $\sqrt{s}$ = 2.76 TeV)} \\ 
          \hline
          Inelastic (INEL) & 0.0705 $\pm$ 0.0007 $\pm$ 0.009 & 0.307$\pm$ 0.003 $\pm$ 0.043 & 0.950 $\pm$ 0.005 $\pm$ 0.026 \\
          \hline   \hline \\
          &  \multicolumn{2}{c}{$\phi$ (pp $\sqrt{s}$ = 2.76 TeV)} \\ 
          \hline
          & d$N$/d$y$  & $\phi$/K$^{-}$  & $\langle{p_{\mathrm{T}}}\rangle$ (GeV/$c$) \\
          \hline
          Inelastic (INEL) & 0.0260 $\pm$ 0.0004 $\pm$ 0.003 & 0.113 $\pm$ 0.001 $\pm$ 0.013 & 1.04 $\pm$ 0.01 $\pm$ 0.09 \\
          \hline  \hline 
        \end{tabular}
      }
      \caption{The values of d$N$/d$y$, ratio to K$^{-}$~\cite{Abelev:2013vea} and 
        $\langle{p_{\mathrm{T}}}\rangle$ are presented for different centrality 
        classes in Pb--Pb collisions and inelastic pp collisions. In each entry, 
        the first uncertainty is statistical and the second is systematic, excluding 
        the normalization uncertainty. Where a third uncertainty is given, it is the 
        normalization uncertainty and the value in the parentheses corresponds to 
        uncorrelated part of the systematic uncertainty.}
      \label{tab_kstar_by_k}
    \end{table} 
    
    \begin{figure}
      \begin{center}
        \includegraphics[scale=0.50]{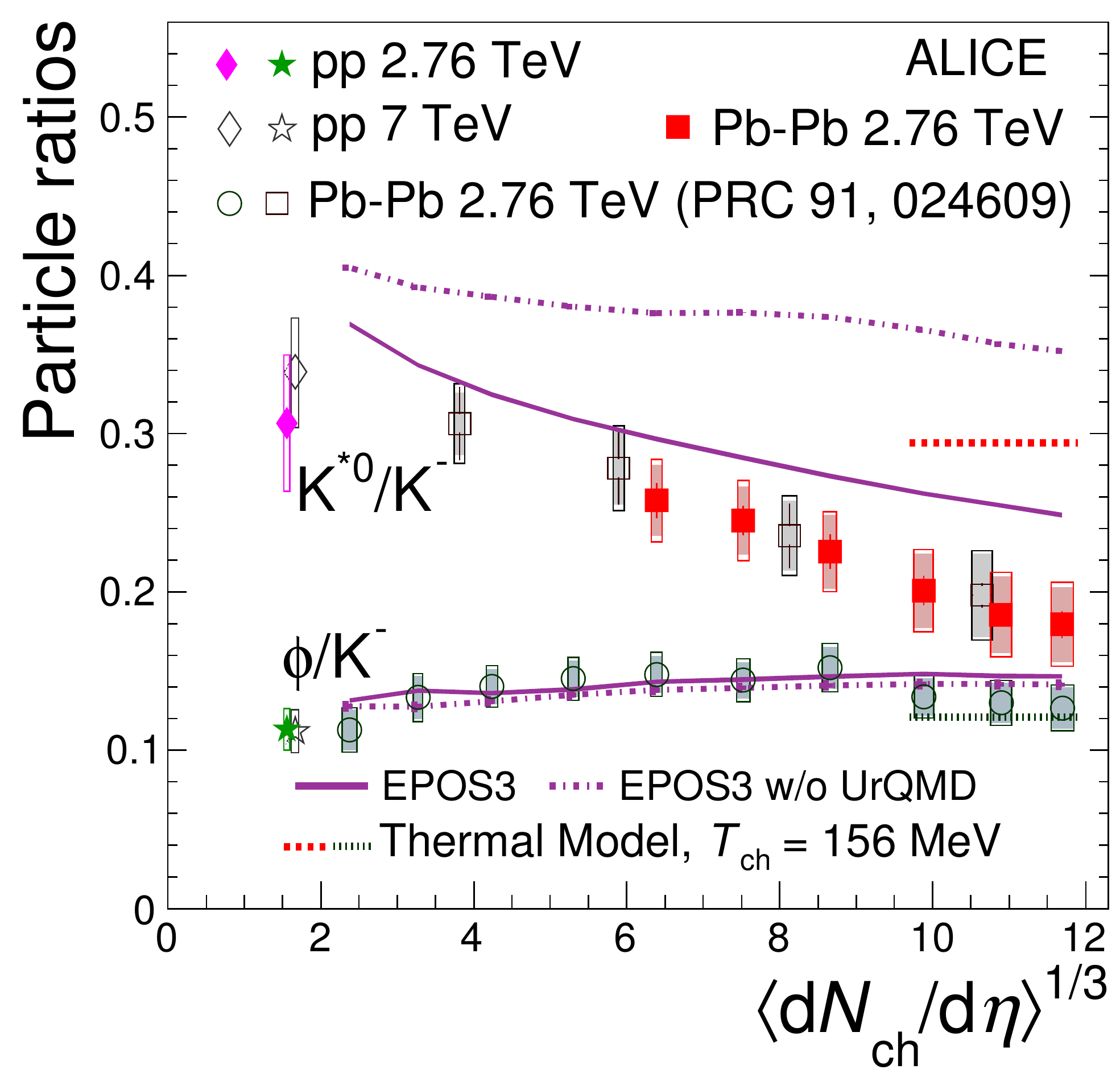}
        \caption{(Color online) K$^{*0}$/K$^{-}$ and $\phi$/K$^{-}$ ratios as a 
          function of $\langle\mathrm{d}N_{\mathrm{ch}}/\mathrm{d}\eta\rangle^{1/3}$
          measured at mid-rapidity~\cite{Aamodt:2011mr} in pp collisions at 
          $\sqrt{s}$ = 2.76 TeV and 7 TeV~\cite{Abelev:2012hy}, and Pb--Pb 
          collisions at $\sqrt{s_\mathrm{NN}}$ = 2.76 TeV. For Pb--Pb collisions, 
          the $\phi$/K$^{-}$ values are exclusively from~\cite{Abelev:2014uua}; 
          the previously published K$^{*0}$/K$^{-}$ measurements are compared to 
          new measurements in finer centrality classes. Bars represent the 
          statistical uncertainties, empty boxes represent the total systematic 
          uncertainties, and shaded boxes represent the systematic uncertainties 
          that are uncorrelated between centrality classes. The expectations from 
          a thermal model calculation with chemical freeze-out temperature of 156 
          MeV for the most central collisions~\cite{Stachel:2013zma} are shown. The 
          EPOS3 calculation of the K$^{*0}$/K and $\phi$/K ratios are also shown 
          as a violet band for different centrality intervals~\cite{Knospe:2015nva}.          
        }
        \label{ratio_kstar_phi}
      \end{center}
    \end{figure}
    
    Figure~\ref{ratio_kstar_phi} shows the ratios  K$^{*0}$/K$^{-}$ and 
    $\phi$/K$^{-}$~\cite{Abelev:2014uua} as a function of 
    $\langle\mathrm{d}N_{\mathrm{ch}}/\mathrm{d}\eta\rangle^{1/3}$ (a proxy for the 
    system size~\cite{Aamodt:2011mr}) in Pb--Pb collisions at $\sqrt{s_\mathrm{NN}}$ 
    = 2.76 TeV and pp collisions at $\sqrt{s}$ = 2.76 TeV and 7 TeV~\cite{Abelev:2012hy}. 
    The yield extraction dominates the systematic uncertainties at low $p_{\mathrm{T}}$, 
    and therefore in the integrated yield; it has been assumed to be fully uncorrelated 
    between different centrality classes. The values of the K$^{*0}$/K$^{-}$ ratio 
    in Pb--Pb collisions at $\sqrt{s_\mathrm{NN}}$ = 2.76 TeV and pp collisions at 
    $\sqrt{s}$ = 2.76 TeV, along with $\phi$/K$^{-}$ ratio in pp collisions at
    $\sqrt{s}$ = 2.76 TeV, are listed in Table~\ref{tab_kstar_by_k}. The K$^{*0}$/K$^{-}$ 
    ratio from the present data is consistent with the trend observed in the previous 
    measurement~\cite{Abelev:2014uua}, also shown in Fig.~\ref{ratio_kstar_phi} for 
    completeness. A smooth dependence on 
    $\langle\mathrm{d}N_{\mathrm{ch}}/\mathrm{d}\eta\rangle^{1/3}$
    is observed and the K$^{*0}$/K$^{-}$ ratio is suppressed in the most central Pb--Pb 
    collisions with respect to pp and peripheral Pb--Pb collisions. On the other hand, 
    the $\phi$/K$^{-}$ ratio (previously reported in~\cite{Abelev:2014uua}) has 
    weak centrality dependence without any suppression. Energy independence of 
    the $\phi$/K$^{-}$ ratio in pp collisions is observed. The suppression of the 
    integrated yield of the short lived K$^{*0}$ resonance suggests that 
    the rescattering of its decay daughters in the hadronic medium reduces 
    the measurable yield of K$^{*0}$. This aspect is further illustrated by 
    comparison of the ratios to a thermal model calculations with a chemical 
    freeze-out temperature of 156 MeV~\cite{Stachel:2013zma}. The measurements of 
    $\phi$/K for the most central collisions agrees with the thermal model 
    expectation, while the measured K$^{*0}$/K ratio lies significantly below 
    the model value as this thermal model does not include rescattering effects.
    The K$^{*0}$/K and $\phi$/K ratios in Pb--Pb collisions are also compared 
    to EPOS3 calculations~\cite{Knospe:2015nva}. EPOS3 is an event generator that 
    describes the full evolution of heavy-ion collisions. The initial conditions 
    are modeled using the Gribov-Regge multiple-scattering framework, based on 
    strings and Pomerons. The collision volume is divided into two parts: a 
    ``core'' (modeled as a QGP described by 3+1 dimensional viscous hydrodynamics) 
    and a ``corona'' (where decaying strings are hadronized). The core is allowed 
    to hadronize and the further evolution of the complete system (including 
    re-scattering and regeneration) is modeled using 
    UrQMD~\cite{Bass:1998ca, Bleicher:1999xi}. EPOS3 with hadronic cascade 
    modeled by UrQMD reproduces the observed trends for K$^{*0}$/K and $\phi$/K 
    ratios in Pb--Pb collisions, suggesting that the observed suppression of
    K$^{*0}$/K ratio is from rescattering of the daughter particles in the 
    hadronic phase.
    
    \begin{figure}[H]
      \begin{center}
        \includegraphics[scale=0.8]{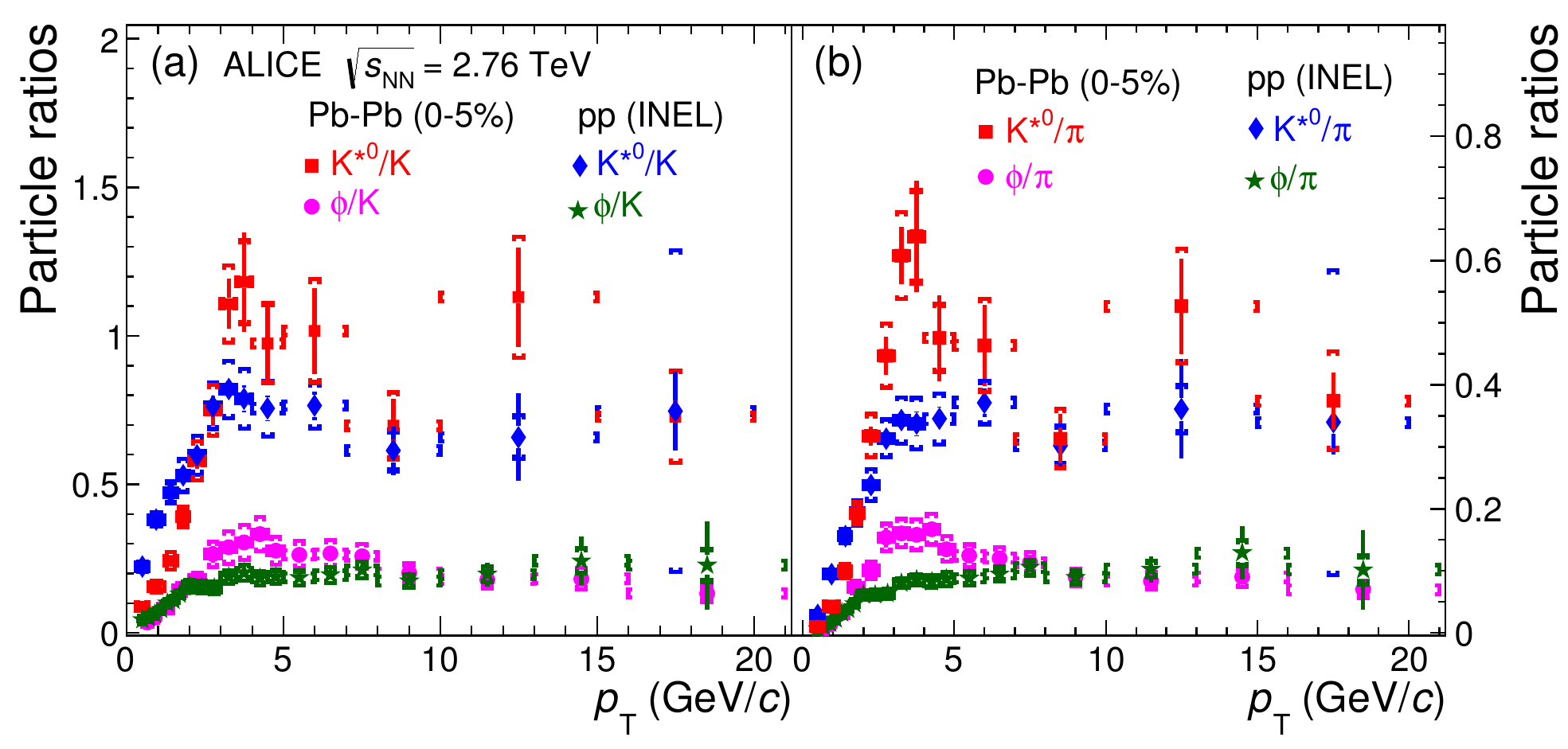}
        \caption{(Color online) Ratios of particle yields K$^{*0}$/K and $\phi/$K in panel 
          (a) and K$^{*0}/\pi$ and $\phi/\pi$ in panel (b) as a function of $p_{\mathrm{T}}$ 
          in central Pb--Pb and pp collisions at $\sqrt{s_\mathrm{NN}}$ = 2.76 TeV 
          are shown. Here, (K$^{*0}$+$\overline{\mathrm{K}}^{*0}$), 
          ($\mathrm{K}^{+}$+$\mathrm{K}^{-}$) and ($\pi^{+}$+$\pi^{-}$) are denoted 
          as K$^{*0}$, K, $\pi$, respectively. The statistical and systematic 
          uncertainties are shown as bars and caps respectively.}
        \label{ratio_phi_k}
      \end{center}
    \end{figure}
    
    The effects of hadronic rescattering can be investigated with the 
    $p_{\mathrm{T}}$-differential K$^{*0}$/K and $\phi$/K ratios. 
    Figure~\ref{ratio_phi_k}a shows the K$^{*0}$/K and $\phi$/K ratios as a 
    function of $p_{\mathrm{T}}$ in pp and 0--5$\%$ central Pb--Pb collisions 
    at $\sqrt{s_\mathrm{NN}}$ = 2.76 TeV. For $p_{\mathrm{T}}$ $\textless$ 
    2 GeV/$c$, the K$^{*0}$/K ratio is smaller in central Pb--Pb collisions than 
    in pp collisions, while the $\phi$/K ratio is the same for both collision
    systems. This is consistent with the suppression of the K$^{*0}$ yield due 
    to rescattering in the hadronic phase. In Fig.~\ref{ratio_phi_k}b, the 
    K$^{*0}/\pi$ and $\phi/\pi$ ratios are shown as a function of $p_{\mathrm{T}}$ 
    in pp and 0--5$\%$ central Pb--Pb collisions at $\sqrt{s_\mathrm{NN}}$ = 2.76 
    TeV. For pp collisions, these ratios saturate at $p_{\mathrm{T}}$ $\sim$4 GeV/$c$, 
    but in Pb--Pb collisions, it increases up to 4 GeV/$c$ then shows a decreasing 
    trend up to 8 GeV/$c$, finally it saturates. Both ratios in central Pb--Pb 
    collisions show an enhancement with respect to pp collisions at $p_{\mathrm{T}}$ 
    $\sim$3 GeV/$c$. Similar meson-to-meson enhancement has been observed for the
    K/$\pi$ ratio~\cite{Abelev:2014laa}, and is understood in terms of radial flow.
    The ratios K$^{*0}/$K, $\phi/$K, K$^{*0}/\pi$ and $\phi/\pi$ are similar at 
    high $p_{\mathrm{T}}$ ($\textgreater$ 8 GeV/$c$) in Pb--Pb and pp collisions. 
    This suggests that fragmentation is the dominant mechanism of hadron production 
    in this $p_{\mathrm{T}}$ regime. This observation is consistent with our 
    previous measurements of the p/$\pi$ and K/$\pi$ ratios~\cite{Abelev:2014laa}. 
    
\begin{figure}[H]
      \begin{center}
        \includegraphics[scale=0.8]{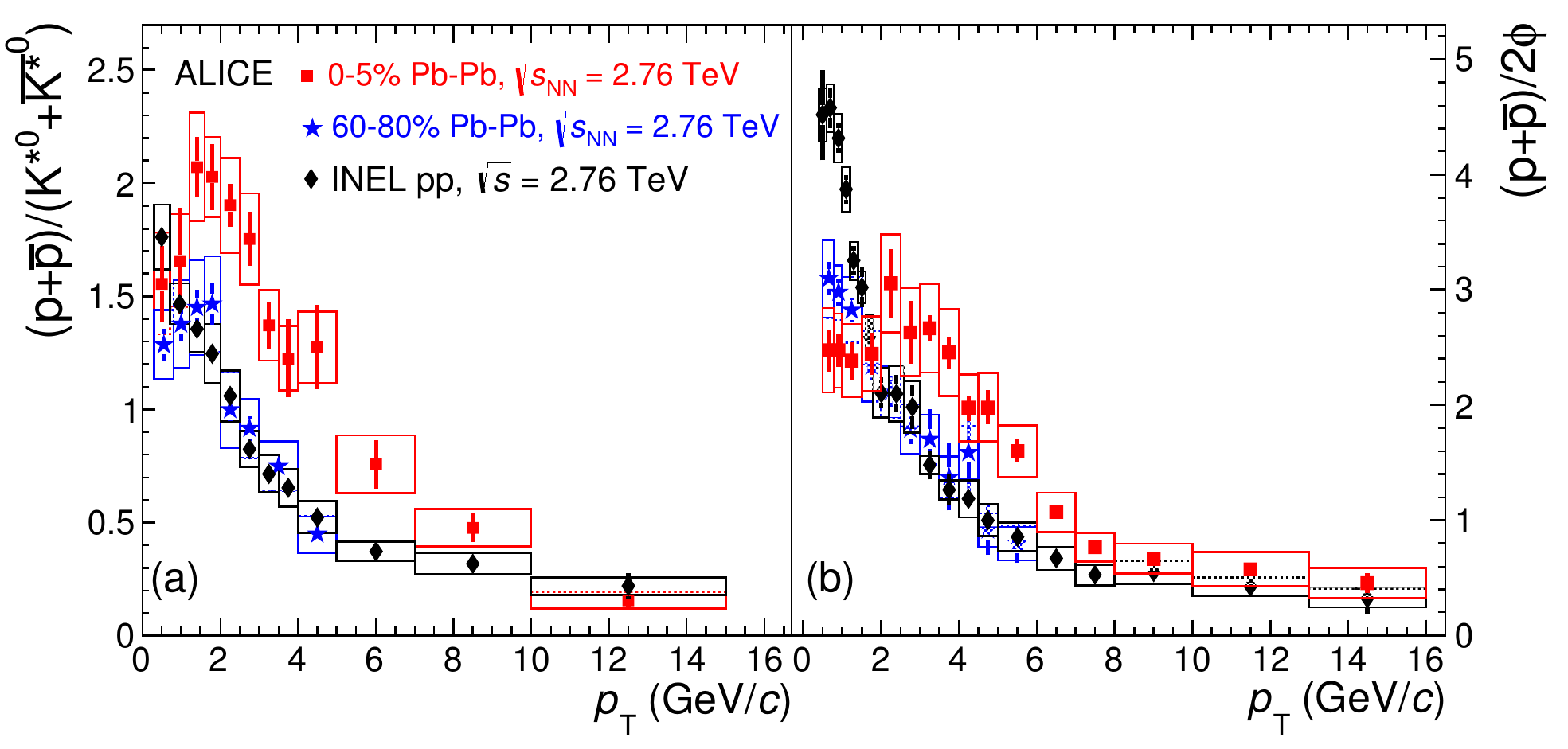}
        \caption{(Color online) Ratios of particle yields p/K$^{*0}$ in panel (a) 
          and p/$\phi$ in panel (b) as a function of $p_{\mathrm{T}}$ in central and 
          peripheral Pb--Pb collisions and pp collisions at $\sqrt{s_\mathrm{NN}}$ = 
          2.76 TeV. The p/$\phi$ ratio for $p_{\mathrm{T}}$ $\textless$ 4 GeV/$c$ is 
          from~\cite{Abelev:2014uua}. The statistical and systematic uncertainties 
          are shown as bars and boxes respectively.}
        \label{ratio_p_kstar_phi}
      \end{center}
    \end{figure}

    In Fig.~\ref{ratio_p_kstar_phi}, the $p_{\mathrm{T}}$-differential p/K$^{*0}$ 
    and p/$\phi$ ratios measured in pp and Pb--Pb collisions at $\sqrt{s_\mathrm{NN}}$ 
    = 2.76 TeV are shown in panels (a) and (b), respectively. The particle ratios
    evolve from pp to central Pb--Pb collisions, indicating a change of the spectral 
    shapes. In central Pb--Pb collisions, the p/K$^{*0}$ ratio shows weak transverse 
    momentum dependence and the p/$\phi$ ratio is consistent with previous observations 
    for $p_{\mathrm{T}}$ $\lesssim$ 4 GeV/$c$. The similarity of the shapes of 
    spectra for K$^{*0}$, p, $\phi$, which have similar masses but different numbers
    of valence quarks, suggests that the shapes are mostly defined by hadron masses 
    as expected from hydrodynamic models~\cite{Shen:2011eg}. At higher $p_{\mathrm{T}}$, 
    the difference between particle ratios measured in different collision systems 
    becomes smaller. Eventually the p/K$^{*0}$ and p/$\phi$ ratios for $p_{\mathrm{T}}$
    $\textgreater$ 8 GeV/$c$ have similar values in both pp and central Pb--Pb 
    collisions within uncertainties as expected if parton fragmentation in vacuum 
    dominates.
  
    \subsection{Nuclear modification factor ($R_\mathrm{AA}$)}\label{sec:raa}
    The $p_{\mathrm{T}}$ spectrum of K$^{*0}$ ($\phi$) in pp collisions is used 
    for the calculation of the nuclear modification factor ($R_\mathrm{AA}$). 
    The K$^{*0}$ spectra is measured up to $p_{\mathrm{T}}$ = 15 GeV/$c$ 
    (Fig.~\ref{pp_pythia_phojet}) and $p_{\mathrm{T}}$ = 20 GeV/$c$ 
    (Fig.~\ref{pbpb_spectrA}), in pp and Pb--Pb collisions, respectively. In pp 
    collisions, the K$^{*0}$ $p_{\mathrm{T}}$ distribution for 15 $\textless$ 
    $p_{\mathrm{T}}$ $\textless$20 GeV/$c$ is extrapolated from the measured data
    using a L\'evy-Tsallis function~\cite{Tsallis:1987eu, Abelev:2006cs}. For the 
    systematic uncertainty on this extrapolated data point, a power-law function is 
    used in the range 2 $\textless$ $p_{\mathrm{T}}$ $\textless$ 20 GeV/$c$. In 
    addition, maximally hard and maximally soft $p_{\mathrm{T}}$ spectra are generated 
    by shifting the measured data points within their uncertainties. The extrapolation 
    procedure is performed on these hard and soft spectra and the changes in the 
    high-$p_{\mathrm{T}}$ yield are incorporated into the systematic uncertainty 
    estimate of the extrapolated data point. 
        
    The $R_\mathrm{AA}$ is used to study the effect of the medium formed in heavy-ion 
    collisions and is sensitive to the system size and the density of the medium. 
    The $R_\mathrm{AA}$ measurement is also sensitive to the dynamics of particle
    production, in-medium effects and the energy loss mechanism of partons in the 
    medium. If a nuclear collision were simply a superposition of nucleon-nucleon 
    collisions, the nuclear modification factor would be equal to unity at 
    high $p_{\mathrm{T}}$. Deviations of $R_\mathrm{AA}$ from unity may indicate the 
    presence of in-medium effects. 

\begin{figure}[H]
      \begin{center}
        \includegraphics[scale=0.7]{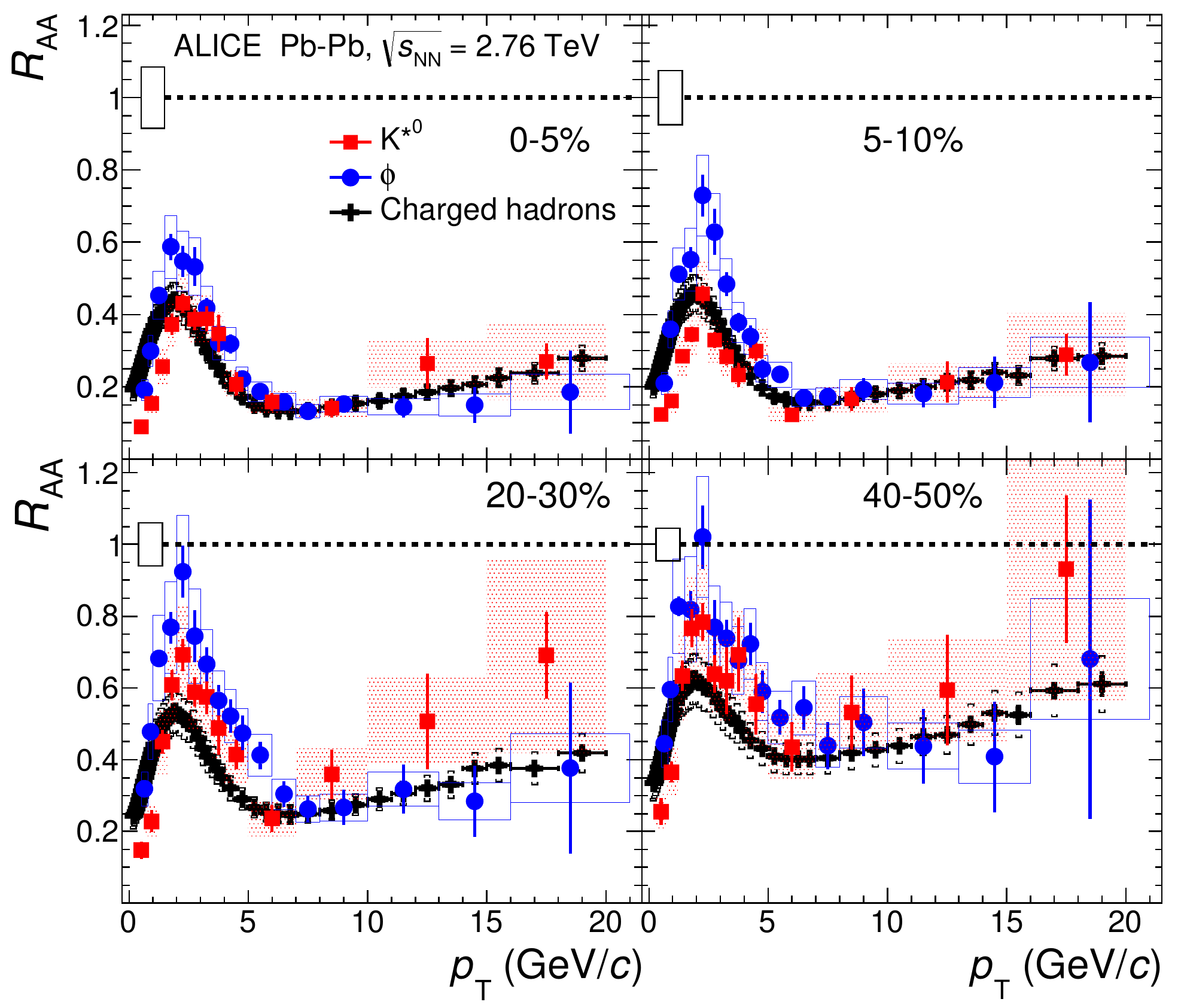}
        \caption{(Color online) The nuclear modification factor, $R_\mathrm{AA}$, as a 
          function of $p_{\mathrm{T}}$ for K$^{*0}$ and $\phi$ mesons in Pb--Pb collisions
          for different centrality classes. The results are compared with the 
          $R_\mathrm{AA}$ of charged hadrons measured by ALICE~\cite{Abelev:2012hxa}. 
          The statistical and systematic uncertainties are shown as bars and boxes, 
          respectively. The boxes around unity indicate the uncertainty on the 
          normalization of $R_\mathrm{AA}$, including the uncertainty on the nuclear 
          overlap function $\langle T_{\mathrm{AA}} \rangle$ and the 
          normalization uncertainty given in Table~\ref{tab_kstar_by_k}.
        }
        \label{raa_1}
      \end{center}
    \end{figure}

    Figure~\ref{raa_1} shows the $R_\mathrm{AA}$ of K$^{*0}$ and $\phi$ in the 0--5$\%$ 
    to 40--50$\%$ centrality classes for Pb--Pb collisions at $\sqrt{s_{\mathrm{NN}}}$ 
    = 2.76 TeV. These results are compared to the $R_\mathrm{AA}$ of charged hadrons 
    measured by the ALICE Collaboration~\cite{Abelev:2012hxa}. The $R_\mathrm{AA}$ of
    K$^{*0}$ and $\phi$ is lower than unity at high $p_\mathrm{T}$ ($\textgreater$ 
    8 GeV/$c$) for all centrality classes. It is also observed that for $p_{\mathrm{T}}$
    $\textless$ 2 GeV/$c$, the K$^{*0}$ $R_\mathrm{AA}$ is smaller than the $\phi$ and
    the charged hadron $R_\mathrm{AA}$. This additional suppression of K$^{*0}$ at 
    low $p_{\mathrm{T}}$ with respect to $\phi$ is reduced as one goes from central 
    to peripheral collisions, consistent with the expectation of more rescattering in 
    central Pb--Pb collisions~\cite{Abelev:2014uua}. At high $p_{\mathrm{T}}$, the 
    $R_\mathrm{AA}$ of both K$^{*0}$ and $\phi$ mesons are similar to that of charged 
    hadrons and the $R_\mathrm{AA}$ values increase from central to peripheral 
    collisions.

\begin{figure}[H]
      \begin{center}
        \includegraphics[scale=0.65]{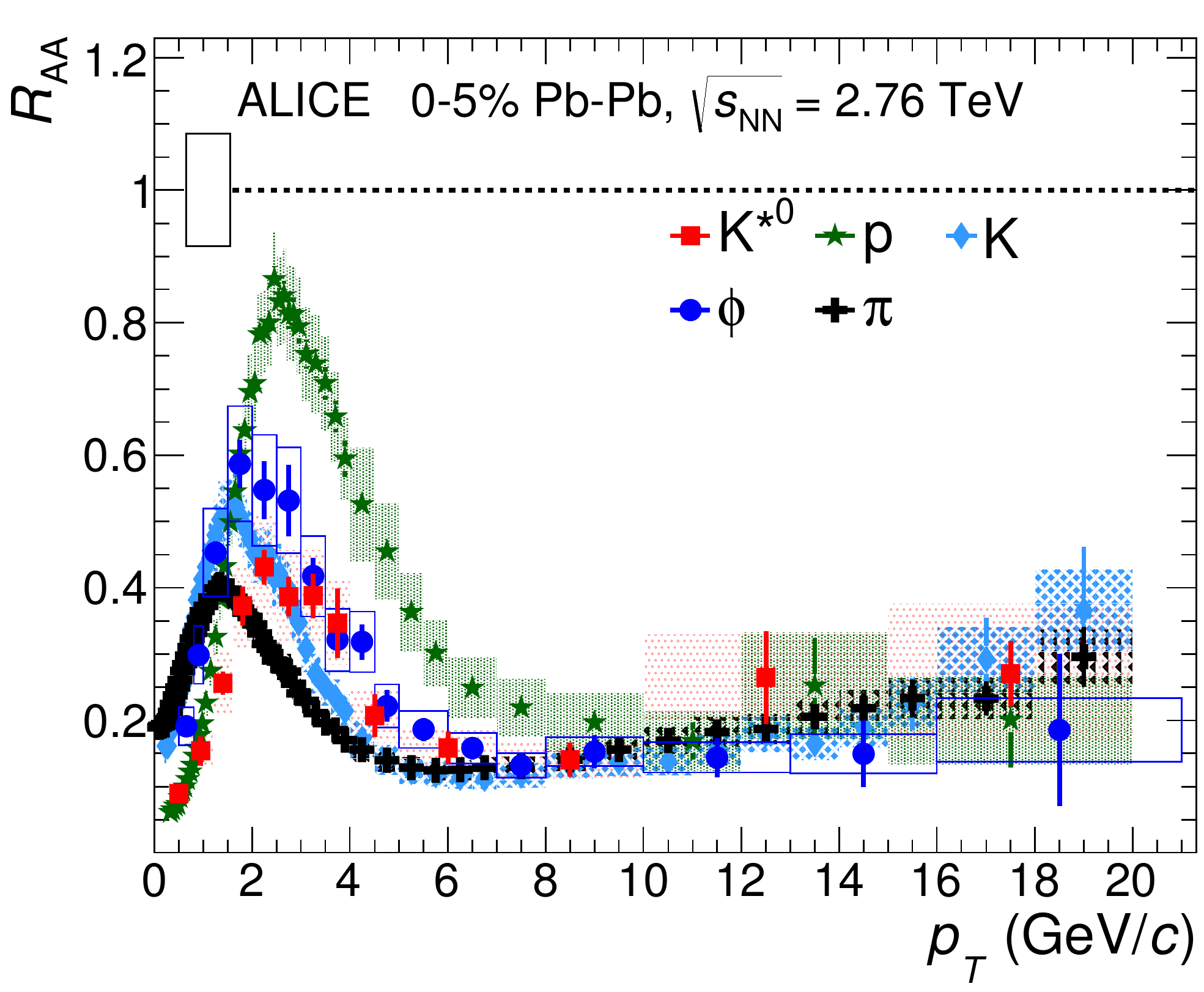}
        \caption{(Color online) The $R_\mathrm{AA}$ for K$^{*0}$ and $\phi$ mesons as a 
          function of $p_{\mathrm{T}}$ in 0--5$\%$ Pb--Pb collisions. The results are 
          compared with the $R_\mathrm{AA}$ of $\pi$, K and p~\cite{Abelev:2014laa}. 
          The statistical and systematic uncertainties are shown as bars and boxes, 
          respectively. The boxes around unity indicate the uncertainty on the 
          normalization of $R_\mathrm{AA}$, including the uncertainty on the nuclear 
          overlap function $\langle T_{\mathrm{AA}} \rangle$ and the normalization 
          uncertainty given in Table~\ref{tab_kstar_by_k}.
        }
        \label{raa_2}
      \end{center}
    \end{figure}

    Figure~\ref{raa_2} shows the comparison of $R_\mathrm{AA}$ of K$^{*0}$ and 
    $\phi$ in the 0--5$\%$ collision centrality class with that of $\pi$, K and 
    p~\cite{Abelev:2014laa}. In the intermediate $p_{\mathrm{T}}$ range (2--6 GeV/$c$), 
    K$^{*0}$ and $\phi$ $R_\mathrm{AA}$ is similar to that of the K, whereas 
    p and $\phi$ exhibit a different trend despite similar masses. The difference
    of $\phi$ and p $R_\mathrm{AA}$ at RHIC was thought to be an effect of 
    hadronization through parton 
    recombination~\cite{Adare:2010pt, Ma:2003ju, Agakishiev:2011dc}. But the 
    p/$\phi$ ratio in most central Pb--Pb collisions at LHC is observed to be 
    flat for $p_{\mathrm{T}}$ $\textless$ 4 GeV/$c$ (see also Fig.~\ref{ratio_p_kstar_phi}b
    and \cite{Abelev:2014uua}) which suggests that particle masses determine the 
    shapes of the $p_{\mathrm{T}}$ spectra with no need to invoke a recombination 
    model. For $p_{\mathrm{T}}$ $\textgreater$ 8 GeV/$c$, all the light flavored 
    species, $\pi$, K, p~\cite{Abelev:2014laa}, K$^{*0}$ and $\phi$, show a similar 
    suppression within uncertainties. This observation rules out models where the 
    suppression of different species containing light quarks are considered to be 
    dependent on their mass and it can also put a stringent constraint on the 
    models dealing with fragmentation and energy loss 
    mechanisms~\cite{Bellwied:2010pr, Liu:2008zb, Liu:2006sf}. 
    
    \section{Conclusions}\label{sec:conc}
    The production of K$^{*0}$ and $\phi$ mesons in inelastic pp collisions and 
    Pb--Pb collisions in various centrality classes at $\sqrt{s_\mathrm{NN}}$ = 2.76 
    TeV using large data samples accumulated in 2011 has been measured. The transverse 
    momentum distributions for K$^{*0}$ ($\phi$) mesons measured in pp collisions up 
    to 15 (21) GeV/$c$  are compared to predictions of the perturbative QCD inspired 
    event generators PYTHIA and PHOJET. It is observed that for $p_{\mathrm{T}}$ 
    $\textgreater$ 8 GeV/$c$ the models agree with the data within uncertainties. In 
    Pb--Pb collisions previously published results for K$^{*0}$ and 
    $\phi$~\cite{Abelev:2014uua} are extended from $p_{\mathrm{T}}$ = 5 GeV/$c$ to 
    20 GeV/$c$ and the production of K$^{*0}$ is studied in finer centrality bins. 
    At high transverse momentum ($p_{\mathrm{T}}$ $\textgreater$ 8 GeV/$c$) nuclear 
    modification factors for different light hadrons ($\pi$, K, K$^{*0}$, p and $\phi$) 
    are consistent within uncertainties and particle ratios (K$^{*0}$/$\pi$, K$^{*0}$/K, 
    $\phi$/$\pi$ and $\phi$/K) are similar for pp and Pb--Pb collisions. This indicates a  
    particle species independence of partonic energy loss in the medium for light 
    quark flavors (u, d, s) and points to fragmentation in vacuum as the dominant 
    particle production mechanism in this kinematic regime. The K$^{*0}$/$\pi$, and 
    $\phi/\pi$ ratios show a centrality dependent enhancement at $p_{\mathrm{T}}$
    $\sim$3 GeV/$c$ in Pb--Pb collisions compared to pp collisions. This is similar 
    to the enhancement previously observed in the K/$\pi$ ratio~\cite{Abelev:2014laa} 
    and attributed to the development of collective radial flow. At low momentum, 
    the production of K$^{*0}$ is significantly suppressed in Pb--Pb collisions and 
    the K$^{*0}$/K ratio exhibits suppression at low momentum, which increases with 
    centrality. This observation is consistent with previous measurements by the 
    STAR~\cite{Adams:2004ep, Aggarwal:2010mt} and the ALICE~\cite{Abelev:2014uua}
    Collaborations and EPOS3 calculations~\cite{Knospe:2015nva}, which confirms the
    importance of rescattering in the hadronic phase. 
    
    \newenvironment{acknowledgement}{\relax}{\relax}
    \begin{acknowledgement}
      \section*{Acknowledgements}

The ALICE Collaboration would like to thank all its engineers and technicians for their invaluable contributions to the construction of the experiment and the CERN accelerator teams for the outstanding performance of the LHC complex.
The ALICE Collaboration gratefully acknowledges the resources and support provided by all Grid centres and the Worldwide LHC Computing Grid (WLCG) collaboration.
The ALICE Collaboration acknowledges the following funding agencies for their support in building and running the ALICE detector:
A. I. Alikhanyan National Science Laboratory (Yerevan Physics Institute) Foundation (ANSL), State Committee of Science and World Federation of Scientists (WFS), Armenia;
Austrian Academy of Sciences and Nationalstiftung f\"{u}r Forschung, Technologie und Entwicklung, Austria;
Ministry of Communications and High Technologies, National Nuclear Research Center, Azerbaijan;
Conselho Nacional de Desenvolvimento Cient\'{\i}fico e Tecnol\'{o}gico (CNPq), Universidade Federal do Rio Grande do Sul (UFRGS), Financiadora de Estudos e Projetos (Finep) and Funda\c{c}\~{a}o de Amparo \`{a} Pesquisa do Estado de S\~{a}o Paulo (FAPESP), Brazil;
Ministry of Science \& Technology of China (MSTC), National Natural Science Foundation of China (NSFC) and Ministry of Education of China (MOEC) , China;
Ministry of Science, Education and Sport and Croatian Science Foundation, Croatia;
Ministry of Education, Youth and Sports of the Czech Republic, Czech Republic;
The Danish Council for Independent Research | Natural Sciences, the Carlsberg Foundation and Danish National Research Foundation (DNRF), Denmark;
Helsinki Institute of Physics (HIP), Finland;
Commissariat \`{a} l'Energie Atomique (CEA) and Institut National de Physique Nucl\'{e}aire et de Physique des Particules (IN2P3) and Centre National de la Recherche Scientifique (CNRS), France;
Bundesministerium f\"{u}r Bildung, Wissenschaft, Forschung und Technologie (BMBF) and GSI Helmholtzzentrum f\"{u}r Schwerionenforschung GmbH, Germany;
Ministry of Education, Research and Religious Affairs, Greece;
National Research, Development and Innovation Office, Hungary;
Department of Atomic Energy Government of India (DAE) and Council of Scientific and Industrial Research (CSIR), New Delhi, India;
Indonesian Institute of Science, Indonesia;
Centro Fermi - Museo Storico della Fisica e Centro Studi e Ricerche Enrico Fermi and Istituto Nazionale di Fisica Nucleare (INFN), Italy;
Institute for Innovative Science and Technology , Nagasaki Institute of Applied Science (IIST), Japan Society for the Promotion of Science (JSPS) KAKENHI and Japanese Ministry of Education, Culture, Sports, Science and Technology (MEXT), Japan;
Consejo Nacional de Ciencia (CONACYT) y Tecnolog\'{i}a, through Fondo de Cooperaci\'{o}n Internacional en Ciencia y Tecnolog\'{i}a (FONCICYT) and Direcci\'{o}n General de Asuntos del Personal Academico (DGAPA), Mexico;
Nationaal instituut voor subatomaire fysica (Nikhef), Netherlands;
The Research Council of Norway, Norway;
Commission on Science and Technology for Sustainable Development in the South (COMSATS), Pakistan;
Pontificia Universidad Cat\'{o}lica del Per\'{u}, Peru;
Ministry of Science and Higher Education and National Science Centre, Poland;
Korea Institute of Science and Technology Information and National Research Foundation of Korea (NRF), Republic of Korea;
Ministry of Education and Scientific Research, Institute of Atomic Physics and Romanian National Agency for Science, Technology and Innovation, Romania;
Joint Institute for Nuclear Research (JINR), Ministry of Education and Science of the Russian Federation and National Research Centre Kurchatov Institute, Russia;
Ministry of Education, Science, Research and Sport of the Slovak Republic, Slovakia;
National Research Foundation of South Africa, South Africa;
Centro de Aplicaciones Tecnol\'{o}gicas y Desarrollo Nuclear (CEADEN), Cubaenerg\'{\i}a, Cuba, Ministerio de Ciencia e Innovacion and Centro de Investigaciones Energ\'{e}ticas, Medioambientales y Tecnol\'{o}gicas (CIEMAT), Spain;
Swedish Research Council (VR) and Knut \& Alice Wallenberg Foundation (KAW), Sweden;
European Organization for Nuclear Research, Switzerland;
National Science and Technology Development Agency (NSDTA), Suranaree University of Technology (SUT) and Office of the Higher Education Commission under NRU project of Thailand, Thailand;
Turkish Atomic Energy Agency (TAEK), Turkey;
National Academy of  Sciences of Ukraine, Ukraine;
Science and Technology Facilities Council (STFC), United Kingdom;
National Science Foundation of the United States of America (NSF) and United States Department of Energy, Office of Nuclear Physics (DOE NP), United States of America.
   
      
    \end{acknowledgement}
    \bibliographystyle{utphys}
    \bibliography{highpT_KStarPhiRAA_v6.4}

\providecommand{\href}[2]{#2}\begingroup\raggedright\begin{thebibliography}{10}

\bibitem{Adams:2005dq}
{\bfseries STAR} Collaboration, J.~Adams {\em et~al.}, ``{Experimental and
  theoretical challenges in the search for the quark gluon plasma: The STAR
  Collaboration's critical assessment of the evidence from RHIC collisions},''
  \href{http://dx.doi.org/10.1016/j.nuclphysa.2005.03.085}{{\em Nucl. Phys.}
  {\bfseries A757} (2005) 102--183},
\href{http://arxiv.org/abs/nucl-ex/0501009}{{\ttfamily arXiv:nucl-ex/0501009
  [nucl-ex]}}.

\bibitem{Adcox:2004mh}
{\bfseries PHENIX} Collaboration, K.~Adcox {\em et~al.}, ``{Formation of dense
  partonic matter in relativistic nucleus-nucleus collisions at RHIC:
  Experimental evaluation by the PHENIX collaboration},''
  \href{http://dx.doi.org/10.1016/j.nuclphysa.2005.03.086}{{\em Nucl. Phys.}
  {\bfseries A757} (2005) 184--283},
\href{http://arxiv.org/abs/nucl-ex/0410003}{{\ttfamily arXiv:nucl-ex/0410003
  [nucl-ex]}}.

\bibitem{Gyulassy:2004zy}
M.~Gyulassy and L.~McLerran, ``{New forms of QCD matter discovered at RHIC},''
  \href{http://dx.doi.org/10.1016/j.nuclphysa.2004.10.034}{{\em Nucl. Phys.}
  {\bfseries A750} (2005) 30--63},
\href{http://arxiv.org/abs/nucl-th/0405013}{{\ttfamily arXiv:nucl-th/0405013
  [nucl-th]}}.

\bibitem{Adare:2006ns}
{\bfseries PHENIX} Collaboration, A.~Adare {\em et~al.}, ``{$J/\psi$ production
  vs centrality, transverse momentum, and rapidity in Au$+$Au collisions at
  $\sqrt{s_{\rm{NN}}}$ = 200 GeV},''
  \href{http://dx.doi.org/10.1103/PhysRevLett.98.232301}{{\em Phys. Rev. Lett.}
  {\bfseries 98} (2007) 232301},
\href{http://arxiv.org/abs/nucl-ex/0611020}{{\ttfamily arXiv:nucl-ex/0611020
  [nucl-ex]}}.

\bibitem{Adare:2011yf}
{\bfseries PHENIX} Collaboration, A.~Adare {\em et~al.}, ``{$J/\psi$
  suppression at forward rapidity in Au$+$Au collisions at $\sqrt{s_{\rm{NN}}}$
  = 200 GeV},'' \href{http://dx.doi.org/10.1103/PhysRevC.84.054912}{{\em Phys.
  Rev.} {\bfseries C84} (2011) 054912},
\href{http://arxiv.org/abs/1103.6269}{{\ttfamily arXiv:1103.6269 [nucl-ex]}}.

\bibitem{Adam:2015nna}
{\bfseries ALICE} Collaboration, J.~Adam {\em et~al.}, ``{Centrality dependence
  of high-$p_\mathrm{T}$ D meson suppression in Pb--Pb collisions at
  $\sqrt{s_{\rm{NN}}}$ = 2.76 TeV},''
  \href{http://dx.doi.org/10.1007/JHEP11(2015)205}{{\em JHEP} {\bfseries 11}
  (2015) 205},
\href{http://arxiv.org/abs/1506.06604}{{\ttfamily arXiv:1506.06604 [nucl-ex]}}.

\bibitem{Adam:2015sza}
{\bfseries ALICE} Collaboration, J.~Adam {\em et~al.}, ``{Transverse momentum
  dependence of D-meson production in Pb--Pb collisions at $\sqrt{s_{\rm{NN}}}$
  = 2.76 TeV},'' \href{http://dx.doi.org/10.1007/JHEP03(2016)081}{{\em JHEP}
  {\bfseries 03} (2016) 081},
\href{http://arxiv.org/abs/1509.06888}{{\ttfamily arXiv:1509.06888 [nucl-ex]}}.

\bibitem{Bellwied:2010pr}
R.~Bellwied and C.~Markert, ``{In-medium hadronization in the deconfined matter
  at RHIC and LHC},''
  \href{http://dx.doi.org/10.1016/j.physletb.2010.06.028}{{\em Phys. Lett.}
  {\bfseries B691} (2010) 208--213},
\href{http://arxiv.org/abs/1005.5416}{{\ttfamily arXiv:1005.5416 [nucl-th]}}.

\bibitem{Liu:2008zb}
W.~Liu and R.~J. Fries, ``{Probing nuclear matter with jet conversions},''
  \href{http://dx.doi.org/10.1103/PhysRevC.77.054902}{{\em Phys. Rev.}
  {\bfseries C77} (2008) 054902},
\href{http://arxiv.org/abs/0801.0453}{{\ttfamily arXiv:0801.0453 [nucl-th]}}.

\bibitem{Liu:2006sf}
W.~Liu, C.~M. Ko, and B.~W. Zhang, ``{Jet conversions in a quark-gluon
  plasma},'' \href{http://dx.doi.org/10.1103/PhysRevC.75.051901}{{\em Phys.
  Rev.} {\bfseries C75} (2007) 051901},
\href{http://arxiv.org/abs/nucl-th/0607047}{{\ttfamily arXiv:nucl-th/0607047
  [nucl-th]}}.

\bibitem{Floris:2014pta}
M.~Floris, ``{Hadron yields and the phase diagram of strongly interacting
  matter},'' \href{http://dx.doi.org/10.1016/j.nuclphysa.2014.09.002}{{\em
  Nucl. Phys.} {\bfseries A931} (2014) 103--112},
\href{http://arxiv.org/abs/1408.6403}{{\ttfamily arXiv:1408.6403 [nucl-ex]}}.

\bibitem{Borsanyi:2010bp}
{\bfseries Wuppertal-Budapest} Collaboration, S.~Bors{\'{a}}nyi, Z.~Fodor,
  C.~Hoelbling, S.~D. Katz, S.~Krieg, C.~Ratti, and K.~K. Szab{\'{o}}, ``{Is
  there still any $T_\mathrm{c}$ mystery in lattice QCD? Results with physical
  masses in the continuum limit III},''
  \href{http://dx.doi.org/10.1007/JHEP09(2010)073}{{\em JHEP} {\bfseries 09}
  (2010) 073},
\href{http://arxiv.org/abs/1005.3508}{{\ttfamily arXiv:1005.3508 [hep-lat]}}.

\bibitem{Aoki:2006br}
Y.~Aoki, Z.~Fodor, S.~D. Katz, and K.~K. Szab{\'{o}}, ``{The QCD transition
  temperature: Results with physical masses in the continuum limit},''
  \href{http://dx.doi.org/10.1016/j.physletb.2006.10.021}{{\em Phys. Lett.}
  {\bfseries B643} (2006) 46--54},
\href{http://arxiv.org/abs/hep-lat/0609068}{{\ttfamily arXiv:hep-lat/0609068
  [hep-lat]}}.

\bibitem{Agashe:2014kda}
{\bfseries Particle Data Group} Collaboration, K.~A. Olive {\em et~al.},
  ``{Review of Particle Physics},''
\href{http://dx.doi.org/10.1088/1674-1137/38/9/090001}{{\em Chin. Phys.}
  {\bfseries C38} (2014) 090001}.

\bibitem{Torrieri:2001ue}
G.~Torrieri and J.~Rafelski, ``{Strange hadron resonances as a signature of
  freezeout dynamics},''
  \href{http://dx.doi.org/10.1016/S0370-2693(01)00492-0}{{\em Phys. Lett.}
  {\bfseries B509} (2001) 239--245},
\href{http://arxiv.org/abs/hep-ph/0103149}{{\ttfamily arXiv:hep-ph/0103149
  [hep-ph]}}.

\bibitem{Adler:2003kg}
{\bfseries PHENIX} Collaboration, S.~S. Adler {\em et~al.}, ``{Scaling
  properties of proton and anti-proton production in $\sqrt{s_{\rm{NN}}}$ = 200
  GeV Au$+$Au collisions},''
  \href{http://dx.doi.org/10.1103/PhysRevLett.91.172301}{{\em Phys. Rev. Lett.}
  {\bfseries 91} (2003) 172301},
\href{http://arxiv.org/abs/nucl-ex/0305036}{{\ttfamily arXiv:nucl-ex/0305036
  [nucl-ex]}}.

\bibitem{Abelev:2013xaa}
{\bfseries ALICE} Collaboration, B.~Abelev {\em et~al.}, ``{$K^0_S$ and
  $\Lambda$ production in Pb--Pb collisions at $\sqrt{s_{\rm{NN}}}$ = 2.76
  TeV},'' \href{http://dx.doi.org/10.1103/PhysRevLett.111.222301}{{\em Phys.
  Rev. Lett.} {\bfseries 111} (2013) 222301},
\href{http://arxiv.org/abs/1307.5530}{{\ttfamily arXiv:1307.5530 [nucl-ex]}}.

\bibitem{Abelev:2014uua}
{\bfseries ALICE} Collaboration, B.~Abelev {\em et~al.}, ``{$K^*(892)^0$ and
  $\phi(1020)$ production in Pb--Pb collisions at $\sqrt{s_{\rm{NN}}}$ = 2.76
  TeV},'' \href{http://dx.doi.org/10.1103/PhysRevC.91.024609}{{\em Phys. Rev.}
  {\bfseries C91} (2015) 024609},
\href{http://arxiv.org/abs/1404.0495}{{\ttfamily arXiv:1404.0495 [nucl-ex]}}.

\bibitem{Miller:2007ri}
M.~L. Miller, K.~Reygers, S.~J. Sanders, and P.~Steinberg, ``{Glauber modeling
  in high energy nuclear collisions},''
  \href{http://dx.doi.org/10.1146/annurev.nucl.57.090506.123020}{{\em Ann. Rev.
  Nucl. Part. Sci.} {\bfseries 57} (2007) 205--243},
\href{http://arxiv.org/abs/nucl-ex/0701025}{{\ttfamily arXiv:nucl-ex/0701025
  [nucl-ex]}}.

\bibitem{Abelev:2013qoq}
{\bfseries ALICE} Collaboration, B.~Abelev {\em et~al.}, ``{Centrality
  determination of Pb--Pb collisions at $\sqrt{s_{\rm{NN}}}$ = 2.76 TeV with
  ALICE},'' \href{http://dx.doi.org/10.1103/PhysRevC.88.044909}{{\em Phys.
  Rev.} {\bfseries C88} no.~4, (2013) 044909},
\href{http://arxiv.org/abs/1301.4361}{{\ttfamily arXiv:1301.4361 [nucl-ex]}}.

\bibitem{Cortese:2004aa}
{\bfseries ALICE} Collaboration, P.~Cortese {\em et~al.}, ``{ALICE technical
  design report on forward detectors: FMD, T0 and V0},''.
\href{http://cds.cern.ch/record/781854?ln=en}{CERN-LHCC-2004-025} (2004).

\bibitem{Abbas:2013taa}
{\bfseries ALICE} Collaboration, E.~Abbas {\em et~al.}, ``{Performance of the
  ALICE VZERO system},''
  \href{http://dx.doi.org/10.1088/1748-0221/8/10/P10016}{{\em JINST} {\bfseries
  8} (2013) P10016},
\href{http://arxiv.org/abs/1306.3130}{{\ttfamily arXiv:1306.3130 [nucl-ex]}}.

\bibitem{Adam:2016xbp}
{\bfseries ALICE} Collaboration, J.~Adam {\em et~al.}, ``{Jet-like correlations
  with neutral pion triggers in pp and central Pb--Pb collisions at 2.76
  TeV},'' \href{http://dx.doi.org/10.1016/j.physletb.2016.10.048}{{\em Phys.
  Lett.} {\bfseries B763} (2016) 238--250},
\href{http://arxiv.org/abs/1608.07201}{{\ttfamily arXiv:1608.07201 [nucl-ex]}}.

\bibitem{Alessandro:2006yt}
{\bfseries ALICE} Collaboration, P.~Cortese {\em et~al.}, ``{ALICE: Physics
  performance report, volume II},''
\href{http://dx.doi.org/10.1088/0954-3899/32/10/001}{{\em J. Phys.} {\bfseries
  G32} (2006) 1295--2040}.

\bibitem{Aamodt:2008zz}
{\bfseries ALICE} Collaboration, K.~Aamodt {\em et~al.}, ``{The ALICE
  experiment at the CERN LHC},''
\href{http://dx.doi.org/10.1088/1748-0221/3/08/S08002}{{\em JINST} {\bfseries
  3} (2008) S08002}.

\bibitem{Abelev:2014ffa}
{\bfseries ALICE} Collaboration, B.~Abelev {\em et~al.}, ``{Performance of the
  ALICE Experiment at the CERN LHC},''
  \href{http://dx.doi.org/10.1142/S0217751X14300440}{{\em Int. J. Mod. Phys.}
  {\bfseries A29} (2014) 1430044},
\href{http://arxiv.org/abs/1402.4476}{{\ttfamily arXiv:1402.4476 [nucl-ex]}}.

\bibitem{Wang:1991hta}
X.-N. Wang and M.~Gyulassy, ``{HIJING: A Monte Carlo model for multiple jet
  production in pp, pA and AA collisions},''
\href{http://dx.doi.org/10.1103/PhysRevD.44.3501}{{\em Phys. Rev.} {\bfseries
  D44} (1991) 3501--3516}.

\bibitem{Sjostrand:2006za}
T.~Sj{\"{o}}strand, S.~Mrenna, and P.~Z. Skands, ``{PYTHIA 6.4 physics and
  manual},'' \href{http://dx.doi.org/10.1088/1126-6708/2006/05/026}{{\em JHEP}
  {\bfseries 05} (2006) 026},
\href{http://arxiv.org/abs/hep-ph/0603175}{{\ttfamily arXiv:hep-ph/0603175
  [hep-ph]}}.

\bibitem{Brun:1994aa}
R.~Brun, F.~Bruyant, F.~Carminati, S.~Giani, M.~Maire, {\em et~al.}, ``{GEANT
  detector description and simulation tool},''.
\href{http://cds.cern.ch/record/1082634}{CERN-W5013} (1994).

\bibitem{Abelev:2012sea}
{\bfseries ALICE} Collaboration, B.~Abelev {\em et~al.}, ``{Measurement of
  inelastic, single- and double-diffraction cross sections in proton--proton
  collisions at the LHC with ALICE},''
  \href{http://dx.doi.org/10.1140/epjc/s10052-013-2456-0}{{\em Eur. Phys. J.}
  {\bfseries C73} no.~6, (2013) 2456},
\href{http://arxiv.org/abs/1208.4968}{{\ttfamily arXiv:1208.4968 [hep-ex]}}.

\bibitem{Abelev:2014laa}
{\bfseries ALICE} Collaboration, B.~Abelev {\em et~al.}, ``{Production of
  charged pions, kaons and protons at large transverse momenta in pp and Pb--Pb
  collisions at $\sqrt{s_{\rm{NN}}}$ = 2.76 TeV},''
  \href{http://dx.doi.org/10.1016/j.physletb.2014.07.011}{{\em Phys. Lett.}
  {\bfseries B736} (2014) 196--207},
\href{http://arxiv.org/abs/1401.1250}{{\ttfamily arXiv:1401.1250 [nucl-ex]}}.

\bibitem{Abelev:2013vea}
{\bfseries ALICE} Collaboration, B.~Abelev {\em et~al.}, ``{Centrality
  dependence of $\pi$, K, p production in Pb--Pb collisions at
  $\sqrt{s_{\rm{NN}}}$ = 2.76 TeV},''
  \href{http://dx.doi.org/10.1103/PhysRevC.88.044910}{{\em Phys. Rev.}
  {\bfseries C88} (2013) 044910},
\href{http://arxiv.org/abs/1303.0737}{{\ttfamily arXiv:1303.0737 [hep-ex]}}.

\bibitem{Sjostrand:2007gs}
T.~Sj{\"{o}}strand, S.~Mrenna, and P.~Z. Skands, ``{A brief introduction to
  PYTHIA 8.1},'' \href{http://dx.doi.org/10.1016/j.cpc.2008.01.036}{{\em
  Comput. Phys. Commun.} {\bfseries 178} (2008) 852--867},
\href{http://arxiv.org/abs/0710.3820}{{\ttfamily arXiv:0710.3820 [hep-ph]}}.

\bibitem{Engel:1994vs}
R.~Engel, ``{Photoproduction within the two component dual parton model:
  Amplitudes and cross-sections},''
\href{http://dx.doi.org/10.1007/BF01496594}{{\em Z. Phys.} {\bfseries C66}
  (1995) 203--214}.

\bibitem{Engel:1995yda}
R.~Engel and J.~Ranft, ``{Hadronic photon-photon interactions at
  high-energies},'' \href{http://dx.doi.org/10.1103/PhysRevD.54.4244}{{\em
  Phys. Rev.} {\bfseries D54} (1996) 4244--4262},
\href{http://arxiv.org/abs/hep-ph/9509373}{{\ttfamily arXiv:hep-ph/9509373
  [hep-ph]}}.

\bibitem{Andersson:1983ia}
B.~Andersson, G.~Gustafson, G.~Ingelman, and T.~Sj{\"{o}}strand, ``{Parton
  fragmentation and string dynamics},''
\href{http://dx.doi.org/10.1016/0370-1573(83)90080-7}{{\em Phys. Rept.}
  {\bfseries 97} (1983) 31--145}.

\bibitem{Field:2008zz}
R.~Field, ``{Physics at the Tevatron},''
{\em Acta Phys. Polon.} {\bfseries B39} (2008) 2611--2672.

\bibitem{Buttar:2004iy}
C.~M. Buttar, D.~Clements, I.~Dawson, and A.~Moraes, ``{Simulations of minimum
  bias events and the underlying event, MC tuning and predictions for the
  LHC},''
{\em Acta Phys. Polon.} {\bfseries B35} (2004) 433--441.

\bibitem{Skands:2010ak}
P.~Z. Skands, ``{Tuning Monte Carlo generators: The Perugia tunes},''
  \href{http://dx.doi.org/10.1103/PhysRevD.82.074018}{{\em Phys. Rev.}
  {\bfseries D82} (2010) 074018},
\href{http://arxiv.org/abs/1005.3457}{{\ttfamily arXiv:1005.3457 [hep-ph]}}.

\bibitem{Abelev:2012hy}
{\bfseries ALICE} Collaboration, B.~Abelev {\em et~al.}, ``{Production of
  $K^*(892)^0$ and $\phi(1020)$ in pp collisions at $\sqrt{s}$ = 7 TeV},''
  \href{http://dx.doi.org/10.1140/epjc/s10052-012-2183-y}{{\em Eur. Phys. J.}
  {\bfseries C72} (2012) 2183},
\href{http://arxiv.org/abs/1208.5717}{{\ttfamily arXiv:1208.5717 [hep-ex]}}.

\bibitem{Tsallis:1987eu}
C.~Tsallis, ``{Possible generalization of Boltzmann-Gibbs statistics},''
\href{http://dx.doi.org/10.1007/BF01016429}{{\em J. Statist. Phys.} {\bfseries
  52} (1988) 479--487}.

\bibitem{Abelev:2006cs}
{\bfseries STAR} Collaboration, B.~I. Abelev {\em et~al.}, ``{Strange particle
  production in p$+$p collisions at $\sqrt{s_{\rm{NN}}}$ = 200 GeV},''
  \href{http://dx.doi.org/10.1103/PhysRevC.75.064901}{{\em Phys. Rev.}
  {\bfseries C75} (2007) 064901},
\href{http://arxiv.org/abs/nucl-ex/0607033}{{\ttfamily arXiv:nucl-ex/0607033
  [nucl-ex]}}.

\bibitem{Schnedermann:1993ws}
E.~Schnedermann, J.~Sollfrank, and U.~W. Heinz, ``{Thermal phenomenology of
  hadrons from 200A GeV S$+$S collisions},''
  \href{http://dx.doi.org/10.1103/PhysRevC.48.2462}{{\em Phys. Rev.} {\bfseries
  C48} (1993) 2462--2475},
\href{http://arxiv.org/abs/nucl-th/9307020}{{\ttfamily arXiv:nucl-th/9307020
  [nucl-th]}}.

\bibitem{Aamodt:2011mr}
{\bfseries ALICE} Collaboration, K.~Aamodt {\em et~al.}, ``{Two-pion
  Bose-Einstein correlations in central Pb--Pb collisions at
  $\sqrt{s_{\rm{NN}}}$ = 2.76 TeV},''
  \href{http://dx.doi.org/10.1016/j.physletb.2010.12.053}{{\em Phys. Lett.}
  {\bfseries B696} (2011) 328--337},
\href{http://arxiv.org/abs/1012.4035}{{\ttfamily arXiv:1012.4035 [nucl-ex]}}.

\bibitem{Stachel:2013zma}
J.~Stachel, A.~Andronic, P.~Braun-Munzinger, and K.~Redlich, ``{Confronting LHC
  data with the statistical hadronization model},''
  \href{http://dx.doi.org/10.1088/1742-6596/509/1/012019}{{\em J. Phys. Conf.
  Ser.} {\bfseries 509} (2014) 012019},
\href{http://arxiv.org/abs/1311.4662}{{\ttfamily arXiv:1311.4662 [nucl-th]}}.

\bibitem{Knospe:2015nva}
A.~G. Knospe, C.~Markert, K.~Werner, J.~Steinheimer, and M.~Bleicher,
  ``{Hadronic resonance production and interaction in partonic and hadronic
  matter in the EPOS3 model with and without the hadronic afterburner UrQMD},''
  \href{http://dx.doi.org/10.1103/PhysRevC.93.014911}{{\em Phys. Rev.}
  {\bfseries C93} no.~1, (2016) 014911},
\href{http://arxiv.org/abs/1509.07895}{{\ttfamily arXiv:1509.07895 [nucl-th]}}.

\bibitem{Bass:1998ca}
S.~A. Bass {\em et~al.}, ``{Microscopic models for ultrarelativistic heavy ion
  collisions},'' \href{http://dx.doi.org/10.1016/S0146-6410(98)00058-1}{{\em
  Prog. Part. Nucl. Phys.} {\bfseries 41} (1998) 255--369},
\href{http://arxiv.org/abs/nucl-th/9803035}{{\ttfamily arXiv:nucl-th/9803035
  [nucl-th]}}.

\bibitem{Bleicher:1999xi}
M.~Bleicher {\em et~al.}, ``{Relativistic hadron hadron collisions in the
  ultrarelativistic quantum molecular dynamics model},''
  \href{http://dx.doi.org/10.1088/0954-3899/25/9/308}{{\em J. Phys.} {\bfseries
  G25} (1999) 1859--1896},
\href{http://arxiv.org/abs/hep-ph/9909407}{{\ttfamily arXiv:hep-ph/9909407
  [hep-ph]}}.

\bibitem{Shen:2011eg}
C.~Shen, U.~Heinz, P.~Huovinen, and H.~Song, ``{Radial and elliptic flow in
  Pb$+$Pb collisions at the Large Hadron Collider from viscous hydrodynamic},''
  \href{http://dx.doi.org/10.1103/PhysRevC.84.044903}{{\em Phys. Rev.}
  {\bfseries C84} (2011) 044903},
\href{http://arxiv.org/abs/1105.3226}{{\ttfamily arXiv:1105.3226 [nucl-th]}}.

\bibitem{Abelev:2012hxa}
{\bfseries ALICE} Collaboration, B.~Abelev {\em et~al.}, ``{Centrality
  dependence of charged particle production at large transverse momentum in
  Pb--Pb collisions at $\sqrt{s_{\rm{NN}}}$ = 2.76 TeV},''
  \href{http://dx.doi.org/10.1016/j.physletb.2013.01.051}{{\em Phys. Lett.}
  {\bfseries B720} (2013) 52--62},
\href{http://arxiv.org/abs/1208.2711}{{\ttfamily arXiv:1208.2711 [hep-ex]}}.

\bibitem{Adare:2010pt}
{\bfseries PHENIX} Collaboration, A.~Adare {\em et~al.}, ``{Nuclear
  modification factors of $\phi$ mesons in $d+$Au, Cu$+$Cu and Au$+$Au
  collisions at $\sqrt{s_{\rm{NN}}}$ = 200 GeV},''
  \href{http://dx.doi.org/10.1103/PhysRevC.83.024909}{{\em Phys. Rev.}
  {\bfseries C83} (2011) 024909},
\href{http://arxiv.org/abs/1004.3532}{{\ttfamily arXiv:1004.3532 [nucl-ex]}}.

\bibitem{Ma:2003ju}
J.~Ma, ``{$\phi$ meson production in $\sqrt{s_{\rm{NN}}}$ = 200 GeV Au$+$Au and
  pp collisions at RHIC},''
  \href{http://dx.doi.org/10.1088/0954-3899/30/1/066}{{\em J. Phys.} {\bfseries
  G30} no.~1, (2004) S543--S548},
\href{http://arxiv.org/abs/nucl-ex/0306014}{{\ttfamily arXiv:nucl-ex/0306014
  [nucl-ex]}}.

\bibitem{Agakishiev:2011dc}
{\bfseries STAR} Collaboration, G.~Agakishiev {\em et~al.}, ``{Identified
  hadron compositions in p$+$p and Au$+$Au collisions at high transverse
  momenta at $\sqrt{s_{\rm{NN}}}$ = 200 GeV},''
  \href{http://dx.doi.org/10.1103/PhysRevLett.108.072302}{{\em Phys. Rev.
  Lett.} {\bfseries 108} (2012) 072302},
\href{http://arxiv.org/abs/1110.0579}{{\ttfamily arXiv:1110.0579 [nucl-ex]}}.

\bibitem{Adams:2004ep}
{\bfseries STAR} Collaboration, J.~Adams {\em et~al.}, ``{$K^*(892)^0$
  resonance production in Au$+$Au and p$+$p collisions at $\sqrt{s_{\rm{NN}}}$
  = 200 GeV at STAR},''
  \href{http://dx.doi.org/10.1103/PhysRevC.71.064902}{{\em Phys. Rev.}
  {\bfseries C71} (2005) 064902},
\href{http://arxiv.org/abs/nucl-ex/0412019}{{\ttfamily arXiv:nucl-ex/0412019
  [nucl-ex]}}.

\bibitem{Aggarwal:2010mt}
{\bfseries STAR} Collaboration, M.~M. Aggarwal {\em et~al.}, ``{$K^{*0}$
  production in Cu$+$Cu and Au$+$Au collisions at $\sqrt{s_{\rm{NN}}}$ = 62.4
  GeV and 200 GeV},'' \href{http://dx.doi.org/10.1103/PhysRevC.84.034909}{{\em
  Phys. Rev.} {\bfseries C84} (2011) 034909},
\href{http://arxiv.org/abs/1006.1961}{{\ttfamily arXiv:1006.1961 [nucl-ex]}}.

\end{thebibliography}\endgroup
    
    \newpage
    \appendix
    
    \section{The ALICE Collaboration}
    \label{app:collab}
    


\begingroup
\small
\begin{flushleft}
J.~Adam$^\textrm{\scriptsize 38}$,
D.~Adamov\'{a}$^\textrm{\scriptsize 87}$,
M.M.~Aggarwal$^\textrm{\scriptsize 91}$,
G.~Aglieri Rinella$^\textrm{\scriptsize 34}$,
M.~Agnello$^\textrm{\scriptsize 30}$\textsuperscript{,}$^\textrm{\scriptsize 113}$,
N.~Agrawal$^\textrm{\scriptsize 47}$,
Z.~Ahammed$^\textrm{\scriptsize 139}$,
S.~Ahmad$^\textrm{\scriptsize 17}$,
S.U.~Ahn$^\textrm{\scriptsize 69}$,
S.~Aiola$^\textrm{\scriptsize 143}$
% ...
\renewcommand\labelenumi{\textsuperscript{\theenumi}~}

\section*{Affiliation notes}
\renewcommand\theenumi{\roman{enumi}}
\begin{Authlist}
\item \Adef{0}Deceased
\item \Adef{idp1816992}{Also at: Georgia State University, Atlanta, Georgia, United States}
\item \Adef{idp3259728}{Also at: Also at Department of Applied Physics, Aligarh Muslim University, Aligarh, India}
\item \Adef{idp4031408}{Also at: M.V. Lomonosov Moscow State University, D.V. Skobeltsyn Institute of Nuclear, Physics, Moscow, Russia}
\end{Authlist}

\section*{Collaboration Institutes}
\renewcommand\theenumi{\arabic{enumi}~}

$^{1}$A.I. Alikhanyan National Science Laboratory (Yerevan Physics Institute) Foundation, Yerevan, Armenia
\\
$^{2}$Benem\'{e}rita Universidad Aut\'{o}noma de Puebla, Puebla, Mexico
\\
$^{3}$Bogolyubov Institute for Theoretical Physics, Kiev, Ukraine
\\
$^{4}$Bose Institute, Department of Physics 
and Centre for Astroparticle Physics and Space Science (CAPSS), Kolkata, India
\\
$^{5}$Budker Institute for Nuclear Physics, Novosibirsk, Russia
\\
$^{6}$California Polytechnic State University, San Luis Obispo, California, United States
\\
$^{7}$Central China Normal University, Wuhan, China
\\
$^{8}$Centre de Calcul de l'IN2P3, Villeurbanne, Lyon, France
\\
$^{9}$Centro de Aplicaciones Tecnol\'{o}gicas y Desarrollo Nuclear (CEADEN), Havana, Cuba
\\
$^{10}$Centro de Investigaciones Energ\'{e}ticas Medioambientales y Tecnol\'{o}gicas (CIEMAT), Madrid, Spain
\\
$^{11}$Centro de Investigaci\'{o}n y de Estudios Avanzados (CINVESTAV), Mexico City and M\'{e}rida, Mexico
\\
$^{12}$Centro Fermi - Museo Storico della Fisica e Centro Studi e Ricerche ``Enrico Fermi', Rome, Italy
\\
$^{13}$Chicago State University, Chicago, Illinois, United States
\\
$^{14}$China Institute of Atomic Energy, Beijing, China
\\
$^{15}$COMSATS Institute of Information Technology (CIIT), Islamabad, Pakistan
\\
$^{16}$Departamento de F\'{\i}sica de Part\'{\i}culas and IGFAE, Universidad de Santiago de Compostela, Santiago de Compostela, Spain
\\
$^{17}$Department of Physics, Aligarh Muslim University, Aligarh, India
\\
$^{18}$Department of Physics, Ohio State University, Columbus, Ohio, United States
\\
$^{19}$Department of Physics, Sejong University, Seoul, South Korea
\\
$^{20}$Department of Physics, University of Oslo, Oslo, Norway
\\
$^{21}$Department of Physics and Technology, University of Bergen, Bergen, Norway
\\
$^{22}$Dipartimento di Fisica dell'Universit\`{a} 'La Sapienza'
and Sezione INFN, Rome, Italy
\\
$^{23}$Dipartimento di Fisica dell'Universit\`{a}
and Sezione INFN, Cagliari, Italy
\\
$^{24}$Dipartimento di Fisica dell'Universit\`{a}
and Sezione INFN, Trieste, Italy
\\
$^{25}$Dipartimento di Fisica dell'Universit\`{a}
and Sezione INFN, Turin, Italy
\\
$^{26}$Dipartimento di Fisica e Astronomia dell'Universit\`{a}
and Sezione INFN, Bologna, Italy
\\
$^{27}$Dipartimento di Fisica e Astronomia dell'Universit\`{a}
and Sezione INFN, Catania, Italy
\\
$^{28}$Dipartimento di Fisica e Astronomia dell'Universit\`{a}
and Sezione INFN, Padova, Italy
\\
$^{29}$Dipartimento di Fisica `E.R.~Caianiello' dell'Universit\`{a}
and Gruppo Collegato INFN, Salerno, Italy
\\
$^{30}$Dipartimento DISAT del Politecnico and Sezione INFN, Turin, Italy
\\
$^{31}$Dipartimento di Scienze e Innovazione Tecnologica dell'Universit\`{a} del Piemonte Orientale and INFN Sezione di Torino, Alessandria, Italy
\\
$^{32}$Dipartimento Interateneo di Fisica `M.~Merlin'
and Sezione INFN, Bari, Italy
\\
$^{33}$Division of Experimental High Energy Physics, University of Lund, Lund, Sweden
\\
$^{34}$European Organization for Nuclear Research (CERN), Geneva, Switzerland
\\
$^{35}$Excellence Cluster Universe, Technische Universit\"{a}t M\"{u}nchen, Munich, Germany
\\
$^{36}$Faculty of Engineering, Bergen University College, Bergen, Norway
\\
$^{37}$Faculty of Mathematics, Physics and Informatics, Comenius University, Bratislava, Slovakia
\\
$^{38}$Faculty of Nuclear Sciences and Physical Engineering, Czech Technical University in Prague, Prague, Czech Republic
\\
$^{39}$Faculty of Science, P.J.~\v{S}af\'{a}rik University, Ko\v{s}ice, Slovakia
\\
$^{40}$Faculty of Technology, Buskerud and Vestfold University College, Tonsberg, Norway
\\
$^{41}$Frankfurt Institute for Advanced Studies, Johann Wolfgang Goethe-Universit\"{a}t Frankfurt, Frankfurt, Germany
\\
$^{42}$Gangneung-Wonju National University, Gangneung, South Korea
\\
$^{43}$Gauhati University, Department of Physics, Guwahati, India
\\
$^{44}$Helmholtz-Institut f\"{u}r Strahlen- und Kernphysik, Rheinische Friedrich-Wilhelms-Universit\"{a}t Bonn, Bonn, Germany
\\
$^{45}$Helsinki Institute of Physics (HIP), Helsinki, Finland
\\
$^{46}$Hiroshima University, Hiroshima, Japan
\\
$^{47}$Indian Institute of Technology Bombay (IIT), Mumbai, India
\\
$^{48}$Indian Institute of Technology Indore, Indore, India
\\
$^{49}$Indonesian Institute of Sciences, Jakarta, Indonesia
\\
$^{50}$Inha University, Incheon, South Korea
\\
$^{51}$Institut de Physique Nucl\'eaire d'Orsay (IPNO), Universit\'e Paris-Sud, CNRS-IN2P3, Orsay, France
\\
$^{52}$Institute for Nuclear Research, Academy of Sciences, Moscow, Russia
\\
$^{53}$Institute for Subatomic Physics of Utrecht University, Utrecht, Netherlands
\\
$^{54}$Institute for Theoretical and Experimental Physics, Moscow, Russia
\\
$^{55}$Institute of Experimental Physics, Slovak Academy of Sciences, Ko\v{s}ice, Slovakia
\\
$^{56}$Institute of Physics, Academy of Sciences of the Czech Republic, Prague, Czech Republic
\\
$^{57}$Institute of Physics, Bhubaneswar, India
\\
$^{58}$Institute of Space Science (ISS), Bucharest, Romania
\\
$^{59}$Institut f\"{u}r Informatik, Johann Wolfgang Goethe-Universit\"{a}t Frankfurt, Frankfurt, Germany
\\
$^{60}$Institut f\"{u}r Kernphysik, Johann Wolfgang Goethe-Universit\"{a}t Frankfurt, Frankfurt, Germany
\\
$^{61}$Institut f\"{u}r Kernphysik, Westf\"{a}lische Wilhelms-Universit\"{a}t M\"{u}nster, M\"{u}nster, Germany
\\
$^{62}$Instituto de Ciencias Nucleares, Universidad Nacional Aut\'{o}noma de M\'{e}xico, Mexico City, Mexico
\\
$^{63}$Instituto de F\'{i}sica, Universidade Federal do Rio Grande do Sul (UFRGS), Porto Alegre, Brazil
\\
$^{64}$Instituto de F\'{\i}sica, Universidad Nacional Aut\'{o}noma de M\'{e}xico, Mexico City, Mexico
\\
$^{65}$IRFU, CEA, Universit\'{e} Paris-Saclay, F-91191 Gif-sur-Yvette, France, Saclay, France
\\
$^{66}$iThemba LABS, National Research Foundation, Somerset West, South Africa
\\
$^{67}$Joint Institute for Nuclear Research (JINR), Dubna, Russia
\\
$^{68}$Konkuk University, Seoul, South Korea
\\
$^{69}$Korea Institute of Science and Technology Information, Daejeon, South Korea
\\
$^{70}$KTO Karatay University, Konya, Turkey
\\
$^{71}$Laboratoire de Physique Corpusculaire (LPC), Clermont Universit\'{e}, Universit\'{e} Blaise Pascal, CNRS--IN2P3, Clermont-Ferrand, France
\\
$^{72}$Laboratoire de Physique Subatomique et de Cosmologie, Universit\'{e} Grenoble-Alpes, CNRS-IN2P3, Grenoble, France
\\
$^{73}$Laboratori Nazionali di Frascati, INFN, Frascati, Italy
\\
$^{74}$Laboratori Nazionali di Legnaro, INFN, Legnaro, Italy
\\
$^{75}$Lawrence Berkeley National Laboratory, Berkeley, California, United States
\\
$^{76}$Moscow Engineering Physics Institute, Moscow, Russia
\\
$^{77}$Nagasaki Institute of Applied Science, Nagasaki, Japan
\\
$^{78}$National and Kapodistrian University of Athens, Physics Department, Athens, Greece, Athens, Greece
\\
$^{79}$National Centre for Nuclear Studies, Warsaw, Poland
\\
$^{80}$National Institute for Physics and Nuclear Engineering, Bucharest, Romania
\\
$^{81}$National Institute of Science Education and Research, Bhubaneswar, India
\\
$^{82}$National Nuclear Research Center, Baku, Azerbaijan
\\
$^{83}$National Research Centre Kurchatov Institute, Moscow, Russia
\\
$^{84}$Niels Bohr Institute, University of Copenhagen, Copenhagen, Denmark
\\
$^{85}$Nikhef, Nationaal instituut voor subatomaire fysica, Amsterdam, Netherlands
\\
$^{86}$Nuclear Physics Group, STFC Daresbury Laboratory, Daresbury, United Kingdom
\\
$^{87}$Nuclear Physics Institute, Academy of Sciences of the Czech Republic, \v{R}e\v{z} u Prahy, Czech Republic
\\
$^{88}$Oak Ridge National Laboratory, Oak Ridge, Tennessee, United States
\\
$^{89}$Petersburg Nuclear Physics Institute, Gatchina, Russia
\\
$^{90}$Physics Department, Creighton University, Omaha, Nebraska, United States
\\
$^{91}$Physics Department, Panjab University, Chandigarh, India
\\
$^{92}$Physics Department, University of Cape Town, Cape Town, South Africa
\\
$^{93}$Physics Department, University of Jammu, Jammu, India
\\
$^{94}$Physics Department, University of Rajasthan, Jaipur, India
\\
$^{95}$Physikalisches Institut, Eberhard Karls Universit\"{a}t T\"{u}bingen, T\"{u}bingen, Germany
\\
$^{96}$Physikalisches Institut, Ruprecht-Karls-Universit\"{a}t Heidelberg, Heidelberg, Germany
\\
$^{97}$Physik Department, Technische Universit\"{a}t M\"{u}nchen, Munich, Germany
\\
$^{98}$Purdue University, West Lafayette, Indiana, United States
\\
$^{99}$Pusan National University, Pusan, South Korea
\\
$^{100}$Research Division and ExtreMe Matter Institute EMMI, GSI Helmholtzzentrum f\"ur Schwerionenforschung GmbH, Darmstadt, Germany
\\
$^{101}$Rudjer Bo\v{s}kovi\'{c} Institute, Zagreb, Croatia
\\
$^{102}$Russian Federal Nuclear Center (VNIIEF), Sarov, Russia
\\
$^{103}$Saha Institute of Nuclear Physics, Kolkata, India
\\
$^{104}$School of Physics and Astronomy, University of Birmingham, Birmingham, United Kingdom
\\
$^{105}$Secci\'{o}n F\'{\i}sica, Departamento de Ciencias, Pontificia Universidad Cat\'{o}lica del Per\'{u}, Lima, Peru
\\
$^{106}$Sezione INFN, Bari, Italy
\\
$^{107}$Sezione INFN, Bologna, Italy
\\
$^{108}$Sezione INFN, Cagliari, Italy
\\
$^{109}$Sezione INFN, Catania, Italy
\\
$^{110}$Sezione INFN, Padova, Italy
\\
$^{111}$Sezione INFN, Rome, Italy
\\
$^{112}$Sezione INFN, Trieste, Italy
\\
$^{113}$Sezione INFN, Turin, Italy
\\
$^{114}$SSC IHEP of NRC Kurchatov institute, Protvino, Russia
\\
$^{115}$Stefan Meyer Institut f\"{u}r Subatomare Physik (SMI), Vienna, Austria
\\
$^{116}$SUBATECH, Ecole des Mines de Nantes, Universit\'{e} de Nantes, CNRS-IN2P3, Nantes, France
\\
$^{117}$Suranaree University of Technology, Nakhon Ratchasima, Thailand
\\
$^{118}$Technical University of Ko\v{s}ice, Ko\v{s}ice, Slovakia
\\
$^{119}$Technical University of Split FESB, Split, Croatia
\\
$^{120}$The Henryk Niewodniczanski Institute of Nuclear Physics, Polish Academy of Sciences, Cracow, Poland
\\
$^{121}$The University of Texas at Austin, Physics Department, Austin, Texas, United States
\\
$^{122}$Universidad Aut\'{o}noma de Sinaloa, Culiac\'{a}n, Mexico
\\
$^{123}$Universidade de S\~{a}o Paulo (USP), S\~{a}o Paulo, Brazil
\\
$^{124}$Universidade Estadual de Campinas (UNICAMP), Campinas, Brazil
\\
$^{125}$Universidade Federal do ABC, Santo Andre, Brazil
\\
$^{126}$University of Houston, Houston, Texas, United States
\\
$^{127}$University of Jyv\"{a}skyl\"{a}, Jyv\"{a}skyl\"{a}, Finland
\\
$^{128}$University of Liverpool, Liverpool, United Kingdom
\\
$^{129}$University of Tennessee, Knoxville, Tennessee, United States
\\
$^{130}$University of the Witwatersrand, Johannesburg, South Africa
\\
$^{131}$University of Tokyo, Tokyo, Japan
\\
$^{132}$University of Tsukuba, Tsukuba, Japan
\\
$^{133}$University of Zagreb, Zagreb, Croatia
\\
$^{134}$Universit\'{e} de Lyon, Universit\'{e} Lyon 1, CNRS/IN2P3, IPN-Lyon, Villeurbanne, Lyon, France
\\
$^{135}$Universit\'{e} de Strasbourg, CNRS, IPHC UMR 7178, F-67000 Strasbourg, France, Strasbourg, France
\\
$^{136}$Universit\`{a} degli Studi di Pavia, Pavia, Italy
\\
$^{137}$Universit\`{a} di Brescia, Brescia, Italy
\\
$^{138}$V.~Fock Institute for Physics, St. Petersburg State University, St. Petersburg, Russia
\\
$^{139}$Variable Energy Cyclotron Centre, Kolkata, India
\\
$^{140}$Warsaw University of Technology, Warsaw, Poland
\\
$^{141}$Wayne State University, Detroit, Michigan, United States
\\
$^{142}$Wigner Research Centre for Physics, Hungarian Academy of Sciences, Budapest, Hungary
\\
$^{143}$Yale University, New Haven, Connecticut, United States
\\
$^{144}$Yonsei University, Seoul, South Korea
\\
$^{145}$Zentrum f\"{u}r Technologietransfer und Telekommunikation (ZTT), Fachhochschule Worms, Worms, Germany
\endgroup

\end{document}